\documentclass{emulateapj}
\usepackage{amsmath}

\def\stacksymbols #1#2#3#4{\def\theguybelow{#2}
        \def\verticalposition{\lower#3pt}
        \def\spacingwithinsymbol{\baselineskip0pt\lineskip#4pt}
        \mathrel{\mathpalette\intermediary#1}}
\def\intermediary #1#2{\verticalposition\vbox{\spacingwithinsymbol
        \everycr={}\tabskip0pt
        \halign{$\mathsurround0pt#1\hfil##\hfil$\crcr#2\crcr
                \theguybelow\crcr}}}

\shorttitle{\emph{Fermi} BUBBLES CREATED WITH AGN JETS}
\shortauthors{GUO \& MATHEWS}

\begin{document}
\bibliographystyle{apj} 

\title {The \emph{Fermi} Bubbles. I. Possible Evidence for Recent AGN Jet Activity in the Galaxy}

\author{Fulai Guo\altaffilmark{1} and William G. Mathews \altaffilmark{1}}

\altaffiltext{1}{UCO/Lick Observatory, Department of Astronomy and Astrophysics, University of California, Santa Cruz, CA 95064, USA; fulai@ucolick.org}

\begin{abstract}

 The {\it Fermi Gamma-ray Space Telescope} reveals two large gamma-ray
 bubbles in the Galaxy, which extend about $50^{\circ}$ ($\sim 10$
 kpc) above and below the Galactic center and are symmetric about the
 Galactic plane. Using axisymmetric hydrodynamic simulations with a
 self-consistent treatment of the dynamical cosmic ray (CR) - gas
 interaction, we show that the bubbles can be created with a recent
 AGN jet activity about $1$ - $3$ Myr ago, which was active for a
 duration of $\sim 0.1$ - $0.5$ Myr. The bipolar jets were ejected
 into the Galactic halo along the rotation axis of the
 Galaxy. Near the Galactic center, the jets must be moderately
 light with a typical density contrast $0.001 \lesssim \eta \lesssim
 0.1$ relative to the ambient hot gas. The jets are energetically
 dominated by kinetic energy, and over-pressured with either CR or
 thermal pressure which induces lateral jet expansion, creating fat CR
 bubbles as observed. The sharp edges of the bubbles imply that CR
 diffusion across the bubble surface is strongly suppressed. The jet
 activity induces a strong shock, which heats and compresses the
 ambient gas in the Galactic halo, potentially explaining the {\it
   ROSAT} X-ray shell features surrounding the bubbles. 
The \emph{Fermi} bubbles provide plausible evidence for a recent powerful AGN jet activity in our Galaxy, 
shedding new insights into the origin of the halo CR population and 
the channel through which massive black holes in disk galaxies release feedback energy during their growth.
\end{abstract}


\section{Introduction}
\label{section:intro}

Using $1.6$ years of data from the {\it Fermi Gamma-ray Space
  Telescope}, \citet{su10} discovered two large gamma-ray bubbles
emitting at $1\lesssim E_{\gamma}\lesssim 100$ GeV in our Galaxy. They
are denoted as the `\emph{Fermi} bubbles', and extend to $\sim 50^{\circ}$
above and below the Galactic center (GC), with a width of about
$40^{\circ}$ in longitude. They have approximately uniform 
gamma ray surface brightness with sharp edges. The bubbles were previously identified as
the `\emph{Fermi} haze' from the first year {\it Fermi} data
\citep{dobler10}. The gamma-ray emission associated with the bubbles
has a hard spectrum ($dN_{\gamma}/dE_{\gamma}\sim E_{\gamma}^{-2}$) and 
could originate from cosmic ray (CR) protons and/or CR electrons. 
CR protons may collide with thermal nuclei inelastically, 
producing neutral pions which decay into gamma rays \citep{crocker11}. 
For the latter, gamma rays can be produced by the inverse Compton (IC) scattering of
CR electrons with the interstellar radiation field (ISRF)
and the cosmic microwave background (CMB; \citealt{dobler10},
\citealt{su10}). The CR electron (CRe) population also produces
microwave synchrotron radiation, which was previously invoked to
explain the microwave haze (at tens of GHz) discovered by the {\it
  Wilkinson Microwave Anisotropy Probe} (WMAP) - the `WMAP haze',
which is spatially associated with the \emph{Fermi} bubbles
(\citealt{finkbeiner04a}; \citealt{dobler08}).

The bilobular morphology of the \emph{Fermi} bubbles suggests that the
responsible CR population is not made of diffused CRs
originally accelerated by supernova (SN) shocks in the Galactic
plane. SN CRs diffusing into the Galactic halo would likely have a
much different spatial distribution, and tend to form one single CR-filled bubble
centered at the GC. Furthermore, the sharp edges of the \emph{Fermi} bubbles imply that
the diffusion is strongly suppressed across the bubble surface, and
thus is not the primary reason for the bubble expansion/formation (see
Section \ref{section:dsuppression}). Similarly, the morphology and
sharp edges of the \emph{Fermi} bubbles suggest that the main CR population in the \emph{Fermi}
bubbles is not byproducts of dark matter annihilations.

As pointed out by \citet{su10}, the \emph{Fermi} bubbles were most likely
created by some large episode of energy injection in the GC, e.g.,
jets originating from an active galactic nucleus (AGN) activity or a
nuclear starburst. Here we perform the first dynamical study of the
jet scenario. Our opposing jet model for the \emph{Fermi} bubbles is inspired
by similar non-thermal double jet events observed in massive galaxies,
particularly those lying at the centers of galaxy groups and clusters
\citep{mcnamara07}. AGN jets
have been previously proposed to explain extended extragalactic
radio sources \citep{longair73,scheuer74,blandford74}.
It is well known that AGN jets carry a significant amount of high energy CRs,
which have been seen from radio synchrotron emission, and 
many AGN jets have also been observed to actively
create CR-filled bubbles (e.g., the jets in Cygnus A and many
Fanaroff-Riley type I radio sources as in \citealt{laing06}). However, due to its limited
sensitivity and resolution, gamma-ray observation can not easily detect AGN bubbles.
The proximity of the \emph{Fermi} bubbles may provide a special opportunity to
thoroughly study AGN bubbles in gamma ray bands.

In this paper we use hydrodynamic simulations to perform a feasibility
study, investigating if the main morphological features of the \emph{Fermi} bubbles 
can be reproduced by a recent AGN jet activity from the GC. Our calculations
take into account the dynamical interactions between CRs and thermal
gas, self-consistently modeling the CR advection with the thermal
gas. Our two-dimensional (2D) calculations allow us to do a fast
parameter study, identifying appropriate jet parameters. We show that
to reproduce the morphology of \emph{Fermi} bubbles, successful AGN jets
cannot be either too massive or too light, but are constrained to have
densities about $0.001$ - $0.1$ times the density of the initial
ambient hot gas near the GC (i.e., with density contrasts
$0.001\lesssim \eta \lesssim 0.1$). In particular, the observed \emph{Fermi}
bubbles can be approximately reproduced by a pair of bipolar AGN jets
which began about $1$ - $3$ Myr ago, and were active for a duration of
$\sim 0.1$ - $0.5$ Myr. The total AGN energy released in our fiducial
run A1 is $\sim 10^{57}$ erg, but it scales with the uncertain density
of the original gas in the Galactic halo, probably varying between about $10^{55}$ and $10^{57}$ erg. 
We also show that to produce the sharp outer edges of
the bubbles seen in gamma ray emission, CR diffusion across the bubble
boundaries must be strongly suppressed below the CR diffusivity observed in 
the solar vicinity .

Similar to other astronomical phenomena, the \emph{Fermi} bubbles may result from 
the interplay of various physical mechanisms. Including hydrodynamics, CR dynamics,
CR advection and diffusion, this paper mainly explores the potentially dominant physics 
involved in the formation of the \emph{Fermi} bubbles -- a pair of opposing jets. 
We follow the dynamical evolution of the jets, and investigate the success and problems of 
the basic jet model by directly comparing with detailed observational features of the \emph{Fermi} bubbles. 
We also investigate the degeneracies and uncertainties associated with the jet model.
In a companion paper (\citealt{guo11c}; hereafter denoted as ``Paper II"),
we investigate the roles of additional physics -- shear viscosity and CR diffusion on the evolution 
of the \emph{Fermi} bubbles,
exploring potential solutions for the observational discrepancies associated with this basic jet scenario. 
The rest of the paper is organized as follows. In
Section~\ref{section2}, we describe CR transport in the Galaxy, our
Galactic model, and numerical setup. We present our results in
Section~\ref{section:results}, and summarize our main conclusions with
implications in Section~\ref{section:conclusion}.
  
\section{The Model and Numerical Methods}
\label{section2}

\subsection{Cosmic Ray Transport in the Galaxy}

The observed \emph{Fermi} bubbles are offset from the GC with
one above and the other below the Galactic plane. They extend to about
$50^{\circ}$ ($\sim 10$ kpc) above and below the GC, and are symmetric
about the Galactic plane. Here we assume that the distance from the
Sun to the GC is $R_{\sun}=8.5$ kpc. In {\it Fermi} maps, the bubbles
are seen at $1\lesssim E_{\gamma}\lesssim 100$ GeV, corresponding to
relativistic CRes with energies $10\lesssim E_{\rm cr}\lesssim 100$
GeV if the observed $\gamma$ rays are inverse Compton upscattering of
starlight by these CRes. The IC cooling time of these CR electrons is
typically a few million years, depending on their energies and
locations (see Figure 28 of \citealt{su10}). Due to the short age and
symmetry of these two bubbles, they probably share the same
origin. The CRs may be accelerated or reaccelerated in situ in these
two bubbles \citep{mertsch11}, but one potential problem for this scenario is to explain
how turbulence is triggered roughly simultaneously in {\it two
  bubble-shaped regions} separated by such a large distance ($\sim 10$
kpc). Here we consider a different scenario where CRs are produced by
one single event and then transported to these two bubbles. If this
scenario is correct, it is likely that the CRs were originated from the
midpoint between the bubbles, the GC.

Each CR particle has a velocity near the speed of light. However, it
is well known that the collective transport speed of CRs in the Galaxy
is much smaller (e.g., CRs reaching the Earth are observed to be
highly isotropic). This is commonly explained by the scattering of CRs
by magnetic irregularities. When the scattering is significant, CRs
are strongly trapped and `effectively' move with magnetic
irregularities, which are frozen in the hot plasma. Therefore, CRs are
advected with thermal gas at the local gas velocity, and in this case,
the CR transport is usually called advection. In regions where the
magnetic field is locally aligned, CRs may stream through thermal gas,
but the streaming speed is limited by the Alfven speed. If CRs stream
along the magnetic field lines at a speed faster than the local Alfven
speed, the CR streaming instability is triggered and excites Alfven
waves, which significantly scatter CRs in return
(\citealt{skilling71}; also see \citealt{kulsrud05},
Chap. 12). Consequently, the CR streaming speed relative to the gas is
roughly the local Alfven speed. Due to the unknown and possibly
complex magnetic field structure, CR streaming relative to the local
gas is often ignored, as we assume in this paper. This is a good
approximation in many cases, where CR scattering off magnetic
irregularities is significant and/or the Alfven speed is much smaller
than the local gas speed.

In addition to advection, CRs can also diffuse through the thermal
gas, as they scatter off magnetic inhomogeneities. CR diffusion is
usually described by the CR diffusion coefficient $\kappa$, which
probably depends on the magnetic field structure and CR
energy. Typical values of $\kappa$ for $1$ GeV CRs are found to be
$\kappa \sim (3-5)\times 10^{28}$ cm$^{2}$ s$^{-1}$ in the solar vicinity \citep{strong07}.

The \emph{Fermi} bubbles extend up to $\sim 10$ kpc away from the GC. If the
CRs in the bubbles are produced at the GC, they must be transported
very quickly, at a speed of $v_{\rm transport}\sim 10 \text{
  kpc}/t_{age}\sim 10000 (t_{\rm age}/1\text{ Myr})^{-1}$ km/s, to
form the bubbles, where $t_{\rm age}$ is
the age of the \emph{Fermi} bubbles. We expect $t_{\rm age}$ to be less than a few million
years due to the short IC cooling times of CRes
responsible for the detected $\gamma$-ray emission. Obviously, the
main CR transport mechanism is not diffusion, which would form one
single gamma-ray bubble centered at the GC and is too slow to
transport CRs. To transport CRs for a distance of $l=10$ kpc within
$t_{\rm age}$ by diffusion, the required diffusion coefficient is
$\kappa \sim l^{2}/t_{\rm age}\sim 3\times 10^{31} (t_{\rm
  age}/1\text{ Myr})^{-1}$ cm$^{2}$ s$^{-1}$, three orders of
magnitude larger than typical estimates of the CR diffusivity in the
Galaxy. Furthermore, diffusion tends to produce blurred boundaries across which 
the CR density varies gradually, 
while the observed \emph{Fermi} bubbles have very sharp edges in gamma rays.

Since diffusion is not responsible for transporting CRs from the GC to
the \emph{Fermi} bubbles, advection of CRs with thermal gas is the only possible CR
transport mechanism. Strong CR advection may be induced by galactic
winds or AGN jets from the GC. Galactic winds are often detected through absorption
lines of their cold gaseous components, which typically have speeds of
about $200$-$300$ km s$^{-1}$ \citep{martin05}, more than one
order of magnitude below the required CR transport speed for the \emph{Fermi} bubbles: 
$v_{\rm transport}\sim 10000 (t_{\rm age}/1\text{ Myr})^{-1}$ km/s. 
However, starburst winds usually contain multi-temperature
components and some hot components may have much larger
speeds \citep{strickland09}. Some winds have indeed been observed to have velocities
slightly beyond $1000$ km/s, particularly those probably driven by
radiation or mechanical energy of AGN events (e.g.,
\citealt{rupke11}). A thermally driven starburst wind flowing at
the speed of $v_{\rm transport}$ would require gas with comparable
sound speeds and temperatures $T \sim 4 \times 10^9$K that are
unreasonably high. Alternatively, the wind scenario may be viable if
the gamma-ray emission from the \emph{Fermi} bubbles is mainly contributed by
the decay of neutral pions produced during the hadronic collisions of
CR protons with thermal nuclei (\citealt{crocker11}). One potential challenge for the
wind scenario is that winds often contain low-temperature gas as seen
in H$\alpha$, molecular, and X-ray filamentary structures, which have
not yet been observed in the \emph{Fermi} bubbles.

In this paper we consider an alternative scenario where CRs are
transported by AGN jets, which typically have relativistic or sub-relativistic velocities on parsec and kpc scales,
much faster than galactic winds. The propagation of AGN jets is determined by the speed of hotspots at the
jet extremities, and may be close to $v_{\rm transport}$, depending on
both jet parameters and the ambient gas properties. The primary goal
of our paper is to study with numerical simulations the possibility of
forming the \emph{Fermi} bubbles with CR-carrying AGN jets from the GC.
CRs in AGN jets may be dominated by the electron-proton plasma or electron-positron
plasma. It is unclear if CR protons or electrons dominate the gamma-ray emission
from the \emph{Fermi} bubbles. In this paper we focus on 
the morphology and dynamical evolution of the \emph{Fermi} bubbles, 
leaving studies of the particle content and gamma-ray emission mechanisms to the future.

CR transport in the Galaxy has been studied numerically by some
simulation codes, in particular, the Galactic Propagation (GALPROP)
code \citep{strong98}, which is very detailed, including
three-dimensional distributions of gas, dust, magnetic fields, optical
and FIR photons. However, GALPROP is not able to self-consistently
model CR advection, which is important when significant gas motions
are present. In addition to jet motions, CR pressure gradients can
also produce gas motions, which in turn advect the CRs. Thus,
dynamical interactions between CRs and the background gas should be
taken into account to self-consistently model CR transport,
particularly advection.

We have developed a finite-differencing code to numerically study CR
transport and the dynamical interaction between CRs and hot gas, which
has been successfully used to study X-ray cavities and radio bubbles
in galaxy clusters in a series of papers, e.g., \citet{mathews08a},
\citet{mathews08}, \citet{mathews09}, \citet{guo10a}, \citet{guo10b},
and \citet{mathews10}. More recently, we use our code to study the
evolution of CR-dominated AGN jets in galaxy clusters in
\citet{guo11}. In this paper, we intend to modify our code to study
AGN jets and the formation of \emph{Fermi} bubbles in the Milky Way. In the
rest of this section, we will elaborate the basic equations and
assumptions of our model, and the setup of our numerical procedure.
      
\subsection{Equations and Assumptions}
\label{section:equation}

The dynamical interaction between CRs and thermal gas may be described
by the CR pressure $P_{\rm c}$. In our calculations, we directly
follow the temporal and spatial evolution of the CR energy density
$e_{\rm c}$, which is related to $P_{\rm c}$ through $P_{\rm
  c}=(\gamma_{\rm c}-1)e_{\rm c}$, where $\gamma_{\rm c}=4/3$. The
nature of the relativistic particles with energy density $e_c$ is
unspecified and may be electrons and/or protons with any spectra. 
Of course the equation of state may be somewhat harder if $e_c$ is mainly contributed
by trans-relativistic protons at $\sim 1$ GeV.
CR pressure gradients can accelerate thermal gas,
converting CR energy into the kinetic energy of thermal gas, which may
be further converted to the internal gas energy in shocks or other
compressions. The combined hydrodynamic evolution of thermal gas and
CRs can be described by the following four equations:
\begin{eqnarray}
\frac{d \rho}{d t} + \rho \nabla \cdot {\bf v} = 0,\label{hydro1}
\end{eqnarray}
\begin{eqnarray}
\rho \frac{d {\bf v}}{d t} = -\nabla (P+P_{\rm c})-\rho \nabla \Phi ,\label{hydro2}
\end{eqnarray}
\begin{eqnarray}
\frac{\partial e}{\partial t} +\nabla \cdot(e{\bf v})=-P\nabla \cdot {\bf v}
   \rm{ ,}\label{hydro3}
   \end{eqnarray}
\begin{eqnarray}
\frac{\partial e_{\rm c}}{\partial t} +\nabla \cdot(e_{\rm c}{\bf v})=-P_{\rm c}\nabla \cdot {\bf v}+\nabla \cdot(\kappa\nabla e_{\rm c})
   \rm{ ,}\label{hydro4}   \end{eqnarray}
  \\ \nonumber
\noindent
where $d/dt \equiv \partial/\partial t+{\bf v} \cdot \nabla $ is the
Lagrangian time derivative, $\kappa$ is the CR diffusion coefficient,
and all other variables have their usual meanings. The gas pressure
$P$ is related to the gas internal energy density $e$ via
$P=(\gamma-1)e$, where we assume $\gamma=5/3$.
 
The Galactic potential $\Phi$ is mainly contributed by three
components: the bulge, disk and dark matter halo. We assume that it is
fixed with time and neglect the small contribution from hot halo
gas. Our simple Galactic potential model and initial
conditions for the thermal gas are explained in Section \ref{section:galmodel}. We
ignore radiative cooling of thermal gas, which is unimportant during
the short-duration ($\lesssim 1$-$3$ Myr) of our simulations. The gas
temperature is related to the gas pressure and density via the ideal
gas law:
\begin{eqnarray}
T=\frac{\mu m_{\mu}P}{k_{\rm B} \rho} {\rm ,}
   \end{eqnarray}
where $k_{\rm B}$ is Boltzmann's constant, $m_{\mu}$ is the atomic
mass unit, and $\mu=0.61$ is the molecular weight. To avoid confusion
we denote the gas pressure $P$ as $P_{\rm g}$ in the rest of the
paper.
  
At time $t=0$ we assume that the CR energy density is zero in the
Galaxy, $e_{\rm c}=0$, but at later times CRs enter the Galactic halo
in jets from the GC. Equation \ref{hydro4} describes the evolution of
CR energy density including both advection and diffusion. CR
diffusion depends on the magnetic field structure and may be anisotropic,
but for simplicity, we generally assume that it is isotropic and the CR diffusion coefficient $\kappa$ is a
constant in space and time: $\kappa \sim 3\times 10^{27}$ - $3\times
10^{29}$ cm$^{2}$ s$^{-1}$ (see Table \ref{table1}). We explore how 
the value of  $\kappa$ affects the formation and morphology of \emph{Fermi}
bubbles in Section \ref{section:dsuppression}, where we also consider
a case with a spatially-varying diffusion coefficient. 
$\kappa$ may also depend on CR energy, which may affect the gamma-ray spectrum 
and the evolution of the \emph{Fermi} bubbles. This effect is not studied in this paper.

As discussed in the previous subsection, CRs interact with magnetic
irregularities and Alfv\'{e}n waves, effectively exerting CR pressure
gradients on the thermal gas (equation \ref{hydro2}). Pressure
gradients in the CR component act directly on the gas by means of
magnetic fields frozen into the gas. This is the primary interaction
between CRs and thermal gas. We neglect other more complicated
(probably secondary) interactions, e.g., Coulomb interactions,
hadronic collisions, and hydromagnetic-wave-mediated CR heating, that
all depend on the CR energy spectrum (e.g., \citealt{guo08a}). Since
the jet evolution in our calculation is mainly driven by its kinetic
energy and the injected CR energy, the dynamical effect of magnetic
fields is expected to be moderate and is unlikely to significantly alter our
results (see our estimates in Section \ref{section:formfermi}). More
sophisticated magnetohydrodynamical calculations are necessary in the
future to explore the impact of magnetic fields. We also neglect CR
energy losses from synchrotron and IC emissions. 
Although the cooling is important for TeV electrons, its impact on 
the integrated CR energy density $e_{\rm c}$ is expected to be small, i.e., the main contribution to $e_{\rm c}$
may be low-energy CR electrons (e.g., at $\sim 0.1$ - $100$ GeV) and
possibly CR protons, which have much longer lifetimes. 
If TeV electrons contribute significantly to the gamma-ray emission 
of the \emph{Fermi} bubbles, CR cooling may have an important effect on the temporal 
evolution of the gamma ray emission (both intensity and spectrum), 
which is beyond the scope of the current paper.

\subsection{The Galactic Model}
\label{section:galmodel}

 \begin{figure}
 %  \epsscale{0.75}
\plotone{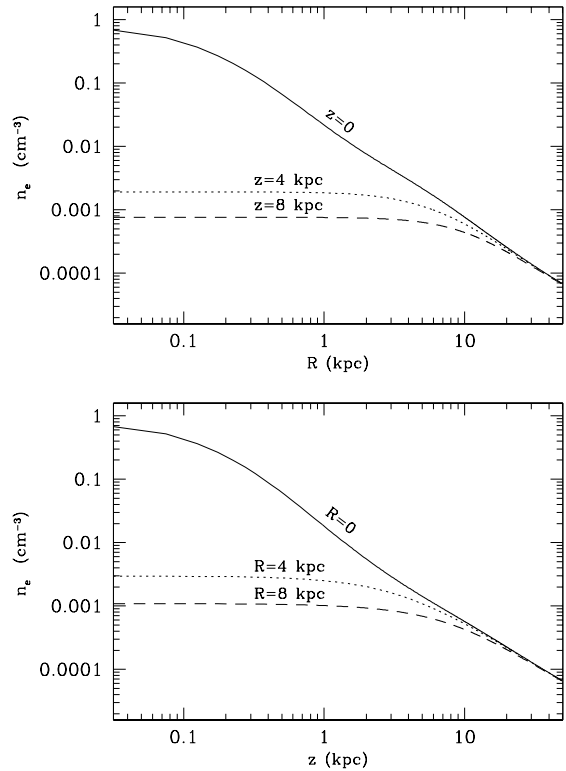}
\caption{Initial profiles of electron number density of the hot gas in
  the Galactic halo along the $R$ direction (parallel to the Galactic
  plane) at three fixed values of $z$ (top panel) and along the $z$
  direction (perpendicular to the Galactic plane) at three fixed
  values of $R$ (bottom panel) in all simulations except run A5, where
  the gas density is lower by a factor of $100$. The density
  distribution $n_{\rm e}(R,z)$ is solved from hydrostatic
  equilibrium, a spatially uniform temperature $T=2.4 \times 10^{6}$
  K, and a given value for the electron number density at the origin,
  $n_{\rm e0}$ (chosen to be $1$ cm$^{-3}$ in all runs except run
  A5).}
 \label{plot1}
 \end{figure}
 
Our key objective in this paper is to perform a feasibility study to
determine if the \emph{Fermi} bubbles can be created with jets from the GC
that are apparently no longer active. The \emph{Fermi} bubbles are inertially
confined by pre-existing gas in the Galactic halo having 
(virial) temperatures
of $\sim 10^{6}$ K. For simplicity, we assume that the hot gaseous
halo is initially isothermal with temperature $T=2.4 \times 10^{6}$ K,
as inferred from X-ray absorption line studies \citep{yao05}. To
derive the initial density distribution of the hot gas, we further
assume that the gas is initially in hydrostatic equilibrium. We adopt
the Galactic potential from \citet{helmi01}, where the Galaxy is
represented by a fixed potential with three components: a dark
logarithmic halo
\begin{eqnarray}
\Phi_{\rm halo}=v_{\rm halo}^{2}\rm{ln}(r^{2}+d_{\rm h}^{2})\rm{,}
\end{eqnarray}
 a Miyamoto-Nagai disk
 \begin{eqnarray}
\Phi_{\rm disc}=-\frac{GM_{\rm disc}}{\sqrt{R^{2}+(a+\sqrt{z^{2}+b^{2}})^{2}}}
\rm{,}
\end{eqnarray} 
 and a spherical Hernquist stellar bulge
 \begin{eqnarray}
\Phi_{\rm bulge}=-\frac{GM_{\rm bulge}}{r+d_{\rm b}}
\rm{,}
\end{eqnarray}  
 where $R$ is the galactocentric radius in the Galactic plane,
 $r=\sqrt{R^{2}+z^{2}}$ is the spatial distance to the GC, $v_{\rm
   halo}=131.5$ km s$^{-1}$ and $d_{\rm h}=12$ kpc; $M_{\rm
   disc}=10^{11} M_{\sun}$, $a=6.5$ kpc and $b=0.26$ kpc; $M_{\rm
   bulge}=3.4\times 10^{10} M_{\sun}$ and $d_{\rm b}=0.7$ kpc. We have
 also explored alternative models of the Galactic potential from
 \citet{b91} and \citet{wolfire95}, and found that the resulting gas
 density distribution is similar.
 
For the hot gas, it is convenient to use the electron number density $n_{\rm e}$, which is 
related to the gas density through
  \begin{eqnarray}
\rho=\mu_{\rm e} n_{\rm e} m_{\mu}
\rm{,}
\end{eqnarray}  
where $\mu_{\rm e}=5\mu/(2+\mu)$ is the molecular weight per
electron. For a given value of electron number density at the origin
$n_{\rm e0}$, we derive the spatial distribution of $n_{\rm e}$ from
hydrostatic equilibrium and the spatially-uniform temperature $T=2.4
\times 10^{6}$ K. In all simulations except run A5, we adopt the
$n_{\rm e}$ distribution with $n_{\rm e0}=1$ cm$^{-3}$, which gives
electron number densities on the order of $10^{-3}$ cm$^{-3}$ at
galactocentric distances of a few kpc. Figure \ref{plot1} shows the
resulting $n_{\rm e}$ profiles along the $R$ direction for three fixed
values of $z$ and along the $z$ direction for three fixed values of
$R$. We have also explored simulations with different values of
$n_{\rm e0}$, and found that, to reproduce the morphology of the
observed \emph{Fermi} bubbles, the required jet power scales with $n_{\rm
  e0}$ (i.e., with the densities of the hot halo gas; see discussions
in Sec. 3.3). In run A5, we choose $n_{\rm e0}=0.01$ cm$^{-3}$,
explicitly exploring the dependence of our results on the density of
halo gas. The unknown density distribution of hot halo gas is a major uncertainty 
in constraining the energetics of the \emph{Fermi} bubble event.
The observed gamma-ray surface brightness can in principle
put a lower limit on the CR energy density within the bubbles, which
may be used to constrain the properties of the hot halo gas 
and the local density of ISRF photons that are IC upscattered.
 
\subsection{Numerical Setup and Jet Injection}
\label{section:jetinjection}

Equations (\ref{hydro1}) $-$ (\ref{hydro4}) are solved in $(R, z)$
cylindrical coordinates using a 2D axisymmetric Eulerian code similar
to ZEUS 2D \citep{stone92}. In particular, our code follows the
evolution of both hot gas and CRs, implementing their dynamical
interaction, CR diffusion, and an energy equation for CRs. The CR
diffusion is solved by using operator-splitting and a fully implicit
Crank-Nicolson differencing method. While this differencing scheme is
stable, we restrict each time step by the stability condition for
explicitly differenced diffusion, as well as Courant conditions for
numerical stability. The computational grid consists of $400$ equally
spaced zones in both coordinates out to $20$ kpc plus additional $100$
logarithmically-spaced zones out to $50$ kpc. The central $20$ kpc in
each direction is resolved to $0.05$ kpc. For both thermal and CR
fluids, we adopt outflow boundary conditions at the outer boundary and
reflective boundary conditions at the inner boundary.

The jet inflow is introduced with a constant speed $v_{\rm jet}$ along
the $z$-axis (the rotation axis of the Galaxy) from the GC. During
the active AGN phase $0\leq t\leq t_{\rm jet}$, the jet is initialized
in a cylinder on the computational grid with radius $R_{\rm jet}$ and
length $z_{\rm jet}$ beginning at the GC. All the physical variables
in this cylinder are updated uniformly with the jet parameters at
every time step during this phase. The jet contains both thermal gas
and relativistic CRs, which are initialized inside the jet source
cylinder and injected into the Galaxy at $z=z_{\rm jet}$ with the jet
speed $v_{\rm jet}$ having an initial opening angle of $0^{\circ}$. In
the simulations presented in this paper, we take $z_{\rm jet}=0.5$
kpc. This allows us to avoid modeling the complex jet evolution closer
to the GC, which would require higher spatial resolution and involves
uncertain multi-phase gas properties. We further assume that the jet
has undergone significant deceleration and transverse expansion within
the jet source region ($0\lesssim z\lesssim z_{\rm jet}$), as seen in
some extragalactic AGN jets (e.g., \citealt{laing06}). In our
calculations, we thus investigate jets with $R_{\rm jet}/z_{\rm
  jet}\sim 0.2-1$ and $v_{\rm jet}/c\sim 0.05-0.3$, where $c$ is the
speed of light. However, it is still possible that AGN jets from the
GC may have much higher velocities at $z_{\rm jet}$. Such jets are
beyond the scope of this paper -- our primary purpose is to perform a
simple feasibility study to investigate if the location, size and morphology
of the observed \emph{Fermi} bubbles can be reproduced by a jet event within
a few Myrs.

We stop the calculation at the current age of the \emph{Fermi} bubbles
$t=t_{\rm Fermi}$, which is defined as the time when the produced CR
bubble reaches the most distant boundary of the observed northern
\emph{Fermi} bubble along the jet axis. The bubble age $t_{\rm Fermi}$ is
constrained to be less than a few million years by the short IC
cooling times of the gamma-ray-emitting CRes within the bubbles. In
this paper we only consider models with $t_{\rm Fermi} \lesssim 3$ Myr
(the estimated IC cooling time of $\sim 100$ GeV electrons at $z\sim
4$ kpc; see Fig. 28 of \citealt{su10}), though it may be dynamically
possible to create the \emph{Fermi} bubbles with a current age more than $3$
Myr. Alternatively, if the gamma-ray emission from the \emph{Fermi} bubbles
is mainly contributed by CR-proton-induced pion decay, the age
of the \emph{Fermi} bubbles could be much longer (see \citealt{crocker11}).

The initial jet is described by six parameters: the thermal gas
density $\rho_{\rm j}$, the gas energy density $e_{\rm j}$, the CR
energy density $e_{ \rm jcr}$, the jet velocity $v_{\rm jet}$, the jet
radius $R_{\rm jet}$, and the jet duration $t_{\rm jet}$. From
$\rho_{\rm j}$, one can derive the commonly-used jet density contrast
between thermal gas inside the jet and the ambient hot gas:
$\eta=\rho_{\rm j}/\rho_{\rm amb}$, where $\rho_{\rm amb}$ is the
initial halo gas density at $(R, z)=(0,z_{\rm jet})$. The jet power
can be written as
\begin{eqnarray}
P_{\rm jet}=P_{\rm ke}+P_{\rm cr} +P_{\rm th}{\rm ,}
   \end{eqnarray}
where $P_{\rm ke}=\rho_{j} v_{\rm jet}^{3}\pi R_{\rm jet}^{2}/2$ is
the jet kinetic power, $P_{\rm cr}=e_{ \rm jcr} v_{\rm jet}\pi R_{\rm
  jet}^{2}$ is the jet CR power, and $P_{\rm th}=e_{ \rm j} v_{\rm
  jet}\pi R_{\rm jet}^{2}$ is the jet thermal power. The total energy
injected by the jet can be written as $E_{\rm jet}=P_{\rm jet}t_{\rm
  jet}$. Thermal and CR energy density are related to the gas and CR
pressure respectively, thus both affecting the dynamical evolution of
hot gas. Therefore $e_{ \rm j}$ and $e_{ \rm jcr}$ are degenerate in
our simulations: we can increase one while decreasing the other,
reaching similar results (see run A6; however, synchrotron and/or IC
emissions from CRs can in principle break this degeneracy). In all
simulation runs except run A6 (see Table \ref{table1}), we take
$\eta_{\rm e}=e_{\rm j}/e_{\rm amb}=1$, where $e_{\rm amb}$ is the
initial ambient gas energy density at $(R, z)=(0,z_{\rm jet})$. In
simulations that successfully produce \emph{Fermi} bubbles as observed, the
jets are usually over-pressured relative to the ambient hot gas at
$z=z_{\rm jet}$. As explained above, the jet simulations have a very
large parameter space, and our 2D calculations permit us to run a
large number (more than $100$) simulations with different parameters
within a reasonable amount of time. Here we present a few
representative calculations particularly relevant to the \emph{Fermi}
bubbles, and list their model parameters in Table \ref{table1}. Some
basic information of the AGN event and the resulting \emph{Fermi} bubbles in
each simulation is given in Table \ref{table2}.

\begin{table*}
 \centering
 \begin{minipage}{130mm}
  \renewcommand{\thefootnote}{\thempfootnote} 
  \caption{Model Parameters in All Simulations Presented in This Paper}
   % \centering
  \begin{tabular}{@{}lccccccccc}
  \hline & {$n_{\rm e0}$\footnote{The initial thermal electron number density at the origin ($R$, $z$)= (0, 0), which determines the density normalization of the isothermal hydrostatic halo gas (see Sec. \ref{section:galmodel}).}}&$\kappa$&$R_{\rm jet}$&$t_{\rm jet}$&{$\eta$\footnote{$\eta$ determines the initial thermal gas density in the jet base $\rho_{\rm j}=\eta \rho_{\rm atm}$, where $\rho_{\rm amb}$ is the initial halo gas density at $(R, z)=(0,z_{\rm jet})$ and $z_{\rm jet}$ is chosen to be $0.5$ kpc in this paper (see Sec. \ref{section:jetinjection}).}}&{$\eta_{\rm e}$\footnote{$\eta_{\rm e}$ determines the initial thermal energy density in the jet base $e_{\rm j}=\eta_{\rm e}e_{\rm atm}$.}}& {$e_{\rm jcr}$\footnote{The initial CR energy density in the jet base (in units of $10^{-9}$ erg cm$^{-3}$).}}&$v_{\rm jet}$& {$n_{\rm ej}$\footnote{$n_{\rm ej}$ is the initial thermal electron number density in the jet base: $n_{\rm ej}=\rho_{\rm j}/(\mu_{\rm e} m_{\mu})$, calculated from the values of $\eta$ and $\rho_{\rm atm}$.}} \\ Run&(cm$^{-3}$)&
        (cm$^{2}$ s$^{-1}$)&(kpc)&(Myr)&&  &&&($10^{-4}$ cm$^{-3}$) \\ \hline 
       A1  .......... &1&$3\times10^{27}$&0.4& 0.3 &0.01&1 &1.0&$0.1c$&5.68 \\ 
      A-diff1  .... &1&$3\times10^{28}$  & 0.4&0.3&0.01&1&1.0&$0.1c$&5.68  \\ 
     A-diff2  .... & 1&$3\times10^{29}$&0.4&  0.3 & 0.01&1 &1.0 &$0.1c$&5.68 \\ 
       A-diff3  .... &1&varied&0.4&0.3 & 0.01&1 &1.0 &$0.1c$&5.68 \\
     A2  ..........  & 1&$3\times10^{27}$&0.4&0.3& 0.02&1 &1.5 &$0.1c$&11.36 \\ 
        A3  ..........  & 1&$3\times10^{27}$&0.4&0.2& 0.01&1 &3.0 &$0.2c$&5.68 \\          
        A4  ..........  & 1&$3\times10^{27}$&0.2&0.3& 0.05&1 &6.0 &$0.1c$&28.40 \\          
       A5  ..........  &0.01&$3\times10^{27}$&0.4& 0.3 &0.01&1 &0.01&$0.1c$&0.057 \\ 
       A6  ..........  &1&$3\times10^{27}$&0.4& 0.3 &0.01&9.325 &0.1&$0.1c$&5.68 \\       
        B1  ..........  & 1&$3\times10^{27}$&0.4&0.3& 0.0001&1 &1.0 &$0.1c$&0.057 \\          
        B2  ..........  &1& $3\times10^{27}$&0.4&0.3& 0.5 &1&1.0 &$0.05c$&284 \\          
          \hline
\label{table1}
\end{tabular}
\end{minipage}
\end{table*}

\begin{table*}
 \centering
 \begin{minipage}{140mm}
  \renewcommand{\thefootnote}{\thempfootnote} 
  \caption{Properties of the \emph{Fermi} Bubbles and the Associated AGN Event in Simulation Runs}
  %  \centering
  \begin{tabular}{@{}lcccccccc}
  \hline &$t_{\rm Fermi}$&{$P_{\rm cr}$\footnote{$P_{\rm cr}$, $P_{\rm ke}$, and $P_{\rm jet}$ are, respectively, the jet CR, kinetic, and total powers (in units of $10^{43}$ erg s$^{-1}$). $P_{\rm jet}=P_{\rm ke}+P_{\rm cr} +P_{\rm th}$, where the thermal jet power $P_{\rm th}$ is much smaller than $P_{\rm ke}$ and/or $P_{\rm cr}$ in our runs. The real jet power depends on the uncertain halo gas density, and is thus not well constrained.}} & {$P_{\rm ke}$\footnotemark}&{$P_{\rm jet}$\footnotemark}&{$E_{\rm jet}$\footnote{$E_{\rm jet}=P_{\rm jet}t_{\rm jet}$ is the energy injected by one jet during the AGN phase $0\leq t\leq t_{\rm jet}$. The total energy injected by both bipolar jets is $2E_{\rm jet}$.}}&{$\dot{M}_{\rm BH}$\footnote{$\dot{M}_{\rm BH}$ is the corresponding accretion rate of the supermassive black hole at the GC, assuming a feedback efficiency of 10\%: $\dot{M}_{\rm BH}=2P_{\rm jet}/(0.1c^{2})$.}} &{$\Delta {M}_{\rm BH}$\footnote{$\Delta {M}_{\rm BH}$ is the total mass accreted by the supermassive black hole at the GC during the AGN event, assuming a feedback efficiency of 10\%: $\Delta {M}_{\rm BH} =\dot{M}_{\rm BH}t_{\rm jet}$.}}&{${M}_{\rm jet}$\footnote{${M}_{\rm jet}=2\rho_{\rm j}\pi R_{\rm jet}^{2}v_{\rm jet}t_{\rm jet}$, estimated at $|z|=0.5$ kpc, is the total mass ejected in the two jets during the AGN event, including the potential gas entrained by the jets at $|z| <0.5$ kpc.}} \\ Run&
        (Myr)& & &($10^{43}$ erg s$^{-1}$)&($10^{56}$ erg)& ($M_{\sun}/$yr)&($M_{\sun}$)&($M_{\sun}$)\\ \hline 
       A1  .............. &2.06&$1.43$&$7.09$&$8.60$&$8.13$&0.03&$9.0\times10^{3}$&$1.5\times10^{5}$\\ 
      A-diff1  ........  &1.94&$1.43$&$7.09$&$8.60$&$8.13$&0.03&$9.0\times10^{3}$&$1.5\times10^{5}$ \\ 
     A-diff2  ........  & 1.30&$1.43$&$7.09$&$8.60$&$8.13$&0.03&$9.0\times10^{3}$&$1.5\times10^{5}$\\ 
       A-diff3  ........  &2.06&$1.43$&$7.09$&$8.60$&$8.13$&0.03&$9.0\times10^{3}$&$1.5\times10^{5}$\\
     A2  ..............  &1.74&$2.15$&$14.18$&$16.41$&$15.51$&0.057&$1.7\times10^{4}$&$3.0\times10^{5}$\\ 
        A3  ..............  &0.86&$8.58$&$56.75$&$65.48$&$41.25$&0.23&$6.9\times10^{4}$&$2.0\times10^{5}$\\          
        A4  ..............  &2.34&$2.15$&$8.87$&$11.03$&$10.42$&0.039&$1.2\times10^{4}$&$1.9\times10^{5}$\\          
        A5  ..............  &2.06&$0.014$&$0.071$&$0.086$&$0.081$&0.0003&90&$1.5\times10^{3}$\\          
        A6  ..............  &2.15&$0.14$&$7.09$&$7.96$&$7.52$&0.028&$8.4\times10^{3}$&$1.5\times10^{5}$\\  
        B1  ..............  &-&$1.43$&$0.07$&$1.58$&$1.49$&0.0055&$1.7\times10^{3}$&$1.5\times10^{3}$\\          
        B2  ..............  & 0.89&$0.72$&$44.33$&$45.08$&$42.6$&0.16&$4.8\times10^{4}$&$3.7\times10^{6}$\\          
          \hline
\label{table2}
\end{tabular}
\end{minipage}
\end{table*}

  \begin{figure*}
    \centering
 %   \epsscale{0.95}
\plottwo {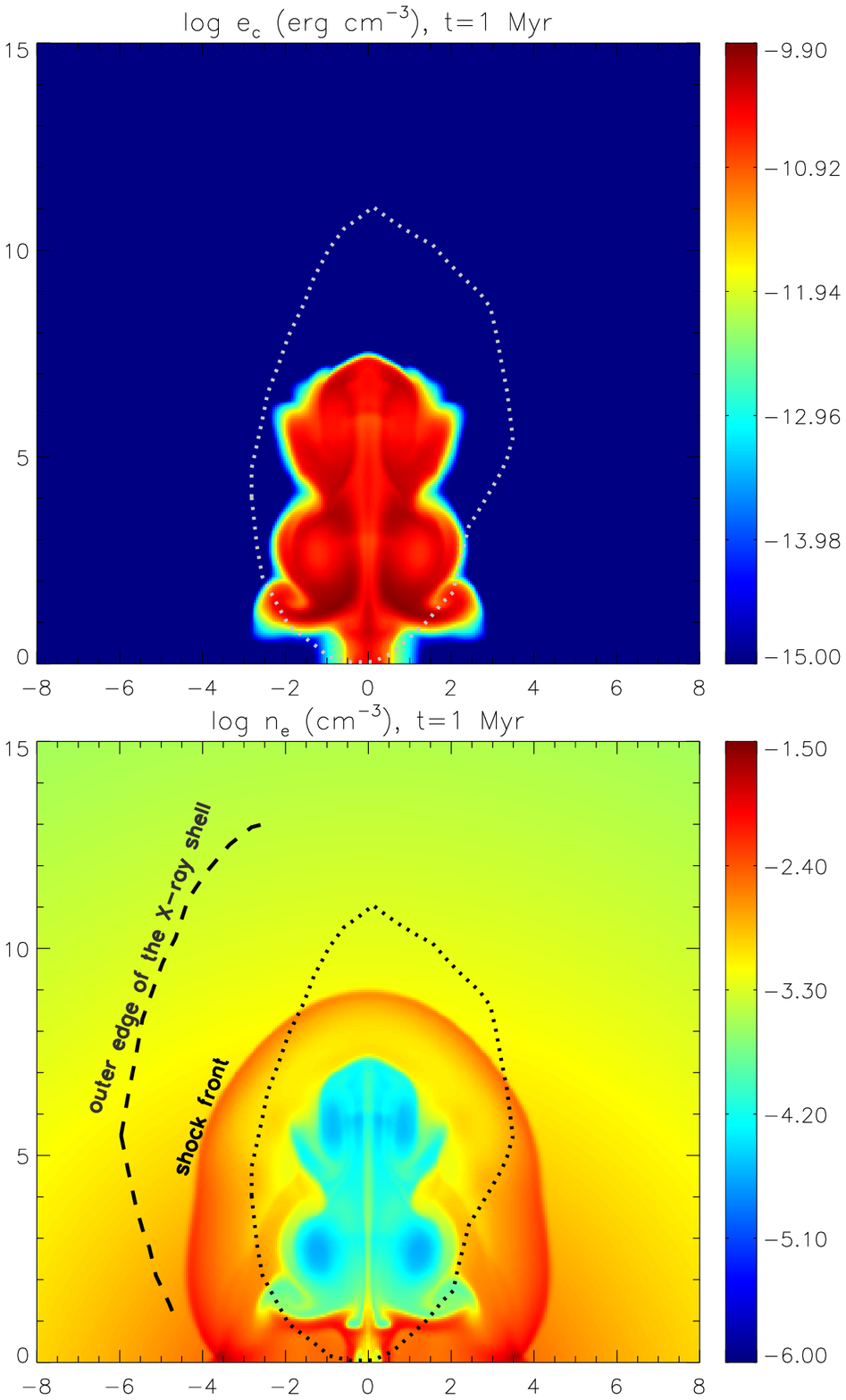} {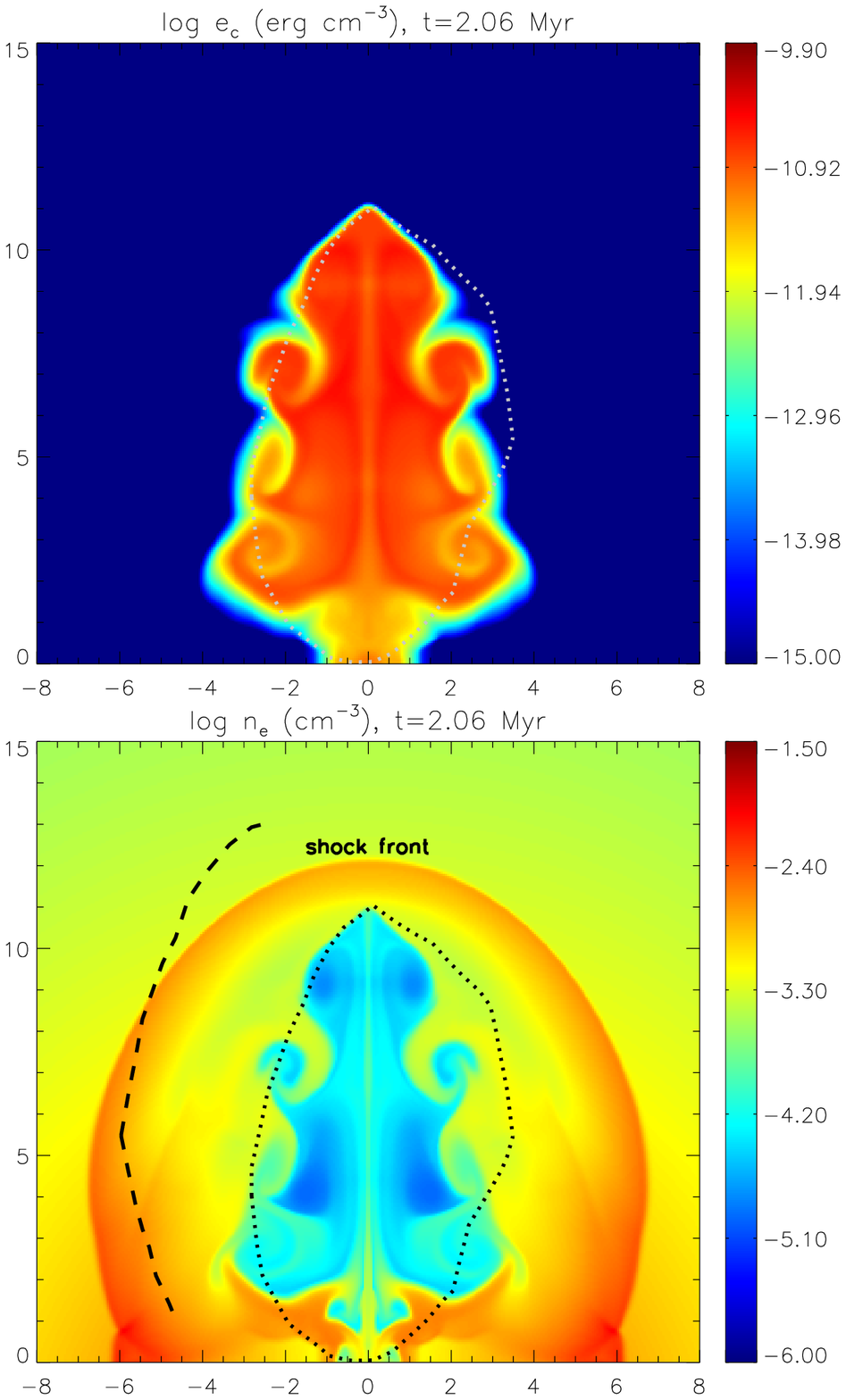} 
\caption{Central slices ($16\times15$ kpc) of CR energy density (top
  panels) and thermal electron number density (bottom panels) in
  logarithmic scale in run A1 at $t=1$ Myr (left panels), and
  $t=t_{\rm Fermi}=2.06$ Myr (right panels). Horizontal and vertical
  axes refer to $R$ and $z$ respectively, labeled in kpc. The dotted
  region in each panel approximately encloses the observed north \emph{Fermi}
  bubble. The propagation of the AGN jet, active for only $t_{\rm
    jet}=0.3$ Myr, produces a CR bubble at $t=2.06$ Myr approximately
  matching the observed \emph{Fermi} bubble. The dashed lines in bottom
  panels trace the outer edge of the {\it ROSAT} X-ray shell
  feature in the northeastern direction (which is most prominent), and is roughly spatially
  coincident with the jet-induced shock at $t=2.06$ Myr.}
 \label{plot2}
 \end{figure*} 
  \begin{figure}
    \centering
 %   \epsscale{0.95}
\plotone {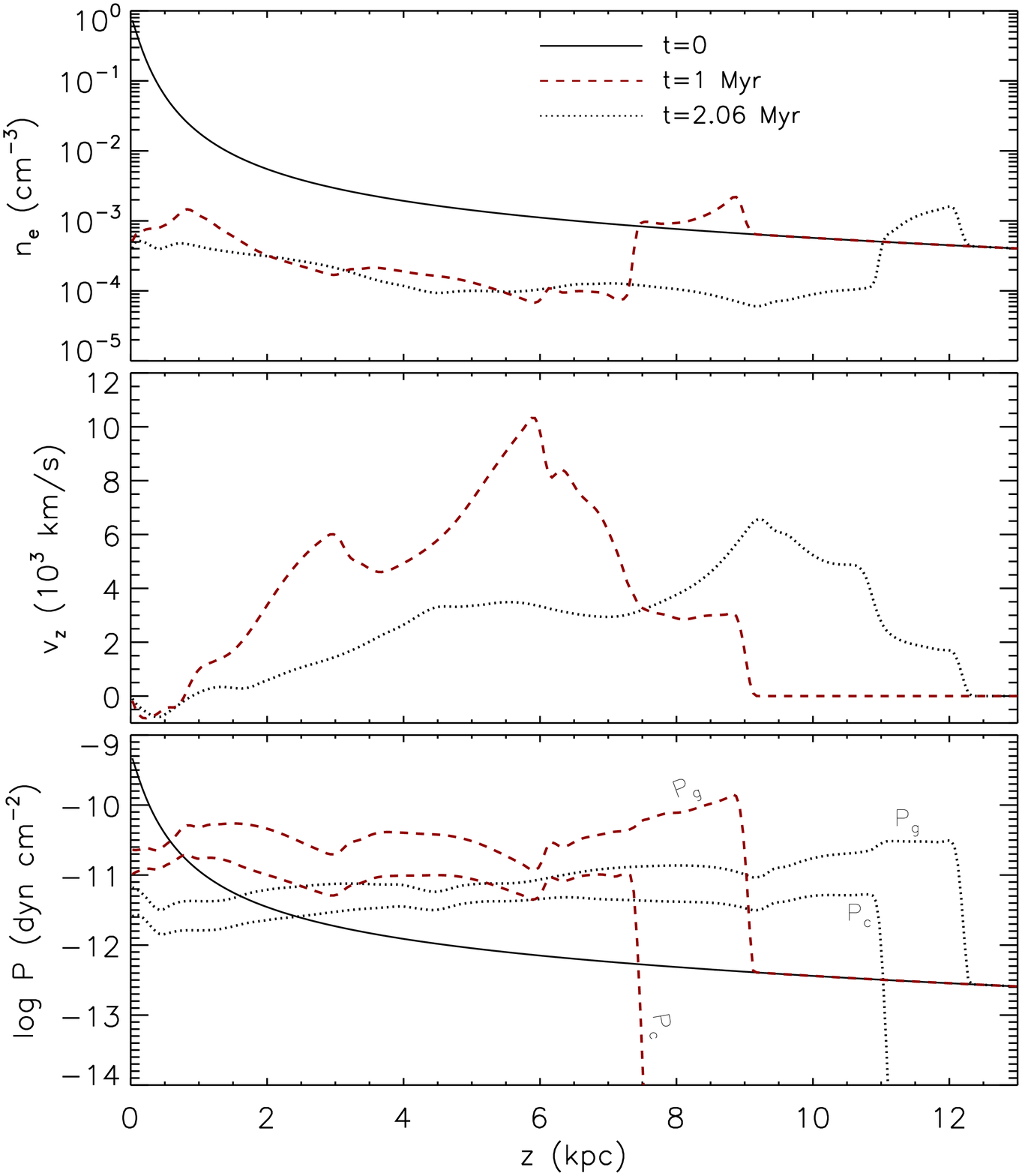} 
\caption{Variations of electron number density (top), the
  $z$-component gas velocity (middle), and pressures (bottom) along
  the jet axis for run A1 at $t=1$ Myr (dashed) and $t=2.06$ Myr
  (dotted). The initial gas density and pressure profiles along the
  $z$-axis at $t=0$ are plotted as solid lines in the top and bottom
  panels, respectively. }
 \label{plot3}
 \end{figure}  
  \begin{figure} 
    \centering
%    \epsscale{0.95}
\plotone {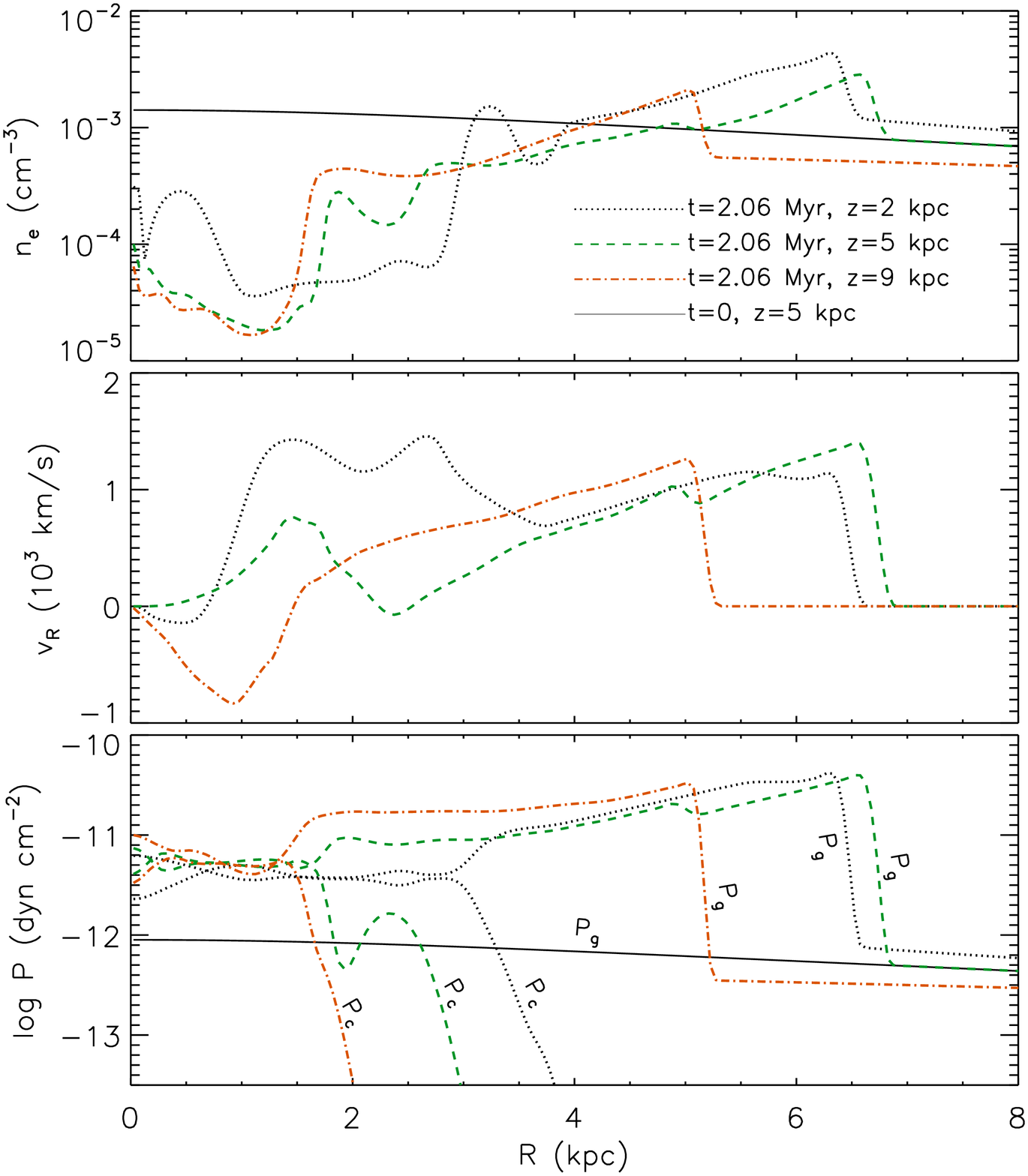} 
\caption{Variations of electron number density (top), the $R$-component gas velocity (middle), and pressures (bottom) along the $R$-direction (perpendicular to the jet axis) for run A1 at $t=2.06$ Myr at $z=2$ (dotted), $5$ (dashed), and $9$ kpc (dot-dashed). The initial gas density and pressure profiles along the $R$-direction at $z=5$ kpc are plotted as solid lines in the top and bottom panels, respectively. }
 \label{plot4}
 \end{figure}
 
 \section{Results}
\label{section:results}

\subsection{Forming \emph{Fermi} Bubbles with AGN Jets}
\label{section:formfermi}

In this subsection, we present in detail one representative run (run
A1), which approximately reproduces the morphology of the \emph{Fermi}
bubbles as observed in gamma-rays with the \emph{Fermi} telescope. The main
objective is to show that it is feasible to produce the observed \emph{Fermi}
bubbles from one single AGN jet activity from the GC. Other choices of
jet parameters may produce similar results, so it is impossible to
determine unique jet properties from the observed bubble
morphology. Here we focus on some interesting features and problems in
this representative run.

The edges of the bubbles are shown in Figure 18 of \citet{su10} in
Galactic coordinates $(l, b)$. To compare our simulation results with
observations, we derive the edges of the bubbles in our coordinates
$(R, z)$ using the following coordinate conversion:  
\begin{eqnarray}
R=R_{\sun}|\text{tan}(l)|
\rm{,}
\end{eqnarray}  
  \begin{eqnarray}
z=\frac{R_{\sun}}{\text{cos}(l)}\text{tan}(b)
\rm{.}
\end{eqnarray}  
The resulting edge of the northern \emph{Fermi} bubble is shown as the dotted
region in each panel of Figure \ref{plot2}. The computed bubble is more
vertically elongated in $(R, z)$ coordinates than 
when viewed in Galactic coordinates since
$\partial z/\partial b\sim 1/\text{cos}^{2}(b)$ increases
significantly with $b$ from $b=0$ to $b\sim 50^{\circ}$. The
line-of-sight projection has a non-negligible effect on the
coordinate conversion, since the bubble size is comparable to the
distance between the Sun and the GC. The real bubble may be slightly
smaller than the one shown in Figure 2. An accurate account for this
effect requires three-dimensional structure of the bubbles, and is
much more complex. We expect this effect to be small and neglect it in
this paper.

The jet parameters for the fiducial run A1 are listed below: radius
$R_{\rm jet}=0.4$ kpc, thermal electron number density $n_{\rm
  ej}=5.68\times 10^{-4}$ cm$^{-3}$, gas energy density $e_{\rm
  j}=5.4\times 10^{-11}$ erg cm$^{-3}$, CR energy density $e_{ \rm
  jcr}=1.0\times 10^{-9}$ erg cm$^{-3}$, velocity $v_{\rm jet}=0.1c$,
and duration $t_{\rm jet}=0.3$ Myr. The jet is light with a density
contrast $\eta=0.01$, which has been previously adopted in many jet
simulations. Due to an additional CR component, the jet is
over-pressured by a factor of $\sim 10$ with regard to the ambient hot
gas at the jet base $z=z_{\rm jet}$. Energetically the kinetic power
in the jet dominates over the CR power by a factor of $\sim 5$ (see
Table \ref{table2}). The simulation is started at $t=0$, and stopped
at the current age of the \emph{Fermi} bubbles $t=t_{\rm Fermi}=2.06$
Myr. The small value of $t_{\rm jet}/t_{\rm Fermi}\sim 0.15$ implies
that all the CRs within the bubbles at the current time have nearly
the same age, consistent with the {\it Fermi} observation that the
gamma-ray spectral index is quite uniform across the whole \emph{Fermi}
bubbles \citep{su10}.

Figure \ref{plot2} shows central slices of CR energy density (top
panels) and thermal electron number density (bottom panels) in
logarithmic scale at $t=1$ Myr (left panels), and $t=t_{\rm Fermi} =
2.06$ Myr (right panels). As clearly seen, the jet produces a rapidly
expanding CR bubble, which roughly matches the current north \emph{Fermi}
bubble (dotted circle) at $t=2.06$ Myr. The CR transport is dominated
by advection due to the high velocity of thermal gas in the jet. The
CR diffusivity in this run is low $\kappa = 3\times 10^{27}$ cm$^{2}$
s$^{-1}$, having a very small effect, which can also be seen by the
sharp edge of the resulting CR bubble (see Section
\ref{section:dsuppression}). The sharp edge is also an important
feature of the observed \emph{Fermi} bubbles. The lateral expansion of the
bubble is mainly due to its high internal pressure (contributed by
both CRs and shocked thermal gas) within the bubble and the rapidly
decreasing ambient gas pressure with the galactocentric distance. The
resulting CR bubble is much narrower if the initial jet has a smaller
internal pressure. Due to the expansion, the gas density in the bubble
is significantly lower than the ambient gas density, as clearly seen
in the bottom panels of Figure \ref{plot2}. This explains why ROSAT
X-ray observations show a `cavity' of X-rays toward the center of the
\emph{Fermi} bubbles (similar to X-ray cavities often found in galaxy
clusters).

The bottom panels of Figure \ref{plot2} show a strong shock produced
by the jet activity. The shock propagates into the hot Galactic halo,
and at $t=t_{\rm Fermi}$ (i.e., the time when the bubbles are
currently being observed), has a speed of $\sim 2000$ km/s (mach
number $M\sim 9$) at the northmost shock front. The propagating shock can
also be seen in Figure \ref{plot3}, which shows variations of thermal
electron number density, the $z$-component of the gas velocity $v_{\rm
  z}$, and pressures along the jet axis at $t=1$ Myr and $t=2.06$
Myr. In directions perpendicular to the jet axis, the shock has a
slightly smaller speed, reaching $R\sim 6$ kpc at $t=t_{\rm Fermi}$,
as seen in Figure \ref{plot2}. Combining with the shock propagating toward the south direction,
the Fermi bubble event produced a dumbbell-shaped shock front.
The shocked gas near the shock front has
higher densities ($n_{\rm e} \sim 1$ - $4\times 10^{-3}$ cm$^{-3}$)
and temperatures ($T\sim 3$ - $7$ keV), potentially explaining the
dumbbell-shaped X-ray shell features detected by the {\it ROSAT} X-ray telescope
(\citealt{sofue00}; \citealt{bland03}). 
The {\it ROSAT} X-ray shell feature is most prominent in the northeastern direction 
(the North Polar Spur; see Fig. 1 in \citealt{sofue00}), and
its outer edge shown as the dashed line in Figure \ref{plot2} 
is roughly spatially coincident with the shock front in the same direction.
At $t=t_{\rm Fermi}$, the shock front in the East direction has a speed of $\sim 1800$ km/s. At
a distance of $R_{\sun}/\text{cos}(35^{\circ})\approx 10.4$ kpc, the
shock front (i.e., the outer edge of the X-ray feature) requires about
$27$ yrs to move an arcsecond, the approximate resolution of the {\it
  Chandra} X-ray Telescope.
Our jet model can not reproduce the bi-conical structure 
in the {\it ROSAT} 1.5 keV map as shown in Figure 5(d) of \citet{bland03}, which
is located within the \emph{Fermi} bubbles and may be produced by a more recent nuclear event.

The strong shock ($M\sim 8$ - $9$) shown in Figures 2, 3 and 4
indicates that the \emph{Fermi} bubbles are expanding explosively, only
slightly decelerated by the small inertial resistance of the
surrounding halo gas. The average total pressure inside our computed
\emph{Fermi} bubbles is about $50$ - $100$ times larger than that in the
un-shocked ambient halo gas. This is very different compared to radio
bubbles and X-ray cavities in galaxy clusters, where usually weak
shocks with Mach number $M\sim 1$ - $2$ are induced. 
Furthermore, buoyancy which is widely recognized to play a key role
in the evolution of X-ray cavities in many galaxy clusters, is much
less important in the evolution of our simulated \emph{Fermi} bubbles, where
the momentum and energy injected by the jets play the dominant role.

The internal structure of the CR bubble can be seen in Figure
\ref{plot4}, which shows variations of electron number density, the
$R$-component gas velocity $v_{\rm R}$, and pressures along the
$R$-direction (perpendicular to the jet axis) at $t=2.06$ Myr at three
different heights: $z=2$ (dotted), $5$ (dashed), and $9$ kpc
(dot-dashed). The bubble corresponds to regions with high CR pressure,
and is surrounded by the outward-propagating shock represented by
large jumps in $n_{\rm e}$, $v_{\rm R}$ and $P_{\rm g}$, which show
the structure of the shock visible in Figure \ref{plot2}. The bubble
has low gas densities $n_{\rm e} \sim 1$ - $ 8 \times10^{-5}$
cm$^{-3}$, and is separated from the surrounding shocked gas through a
contact discontinuity across which the total pressure is
continuous. Within the bubble, the CR pressure in run A1 is typically $\sim 3$ -
$6\times10^{-12}$ dyn cm$^{-2}$, which is comparable to the gas
pressure, and slightly larger than the magnetic pressure of the $5$ -
$10$ $\mu$G fields considered by \citet{su10} to explain the ``WMAP
haze" observed at the base of the \emph{Fermi} bubbles as synchrotron
emission. This supports our neglect of the magnetic field in the
bubble dynamics, which will likely not significantly change our
results. However, as discussed in the last paragraph of Section 3.3,
the CR pressure in our simulated bubbles depends on a model parameter -- the initial
CR pressure at the jet base, and does not represent the real CR pressure in the \emph{Fermi} bubbles, 
which may be constrained by microwave and gamma ray emissions from the bubbles (see Sec. 3.5).

In run A1, the jet is energetically dominated by the kinetic energy,
with the kinetic power $P_{\rm ke}\sim 7.09 \times 10^{43}$ erg
s$^{-1}$ and the CR power $P_{\rm cr}\sim 1.43 \times 10^{43}$ erg
s$^{-1}$. The total jet power is $P_{\rm jet}\sim 8.6 \times 10^{43}$
erg s$^{-1}$. The jet is assumed to be produced by Sgr A*, the supermassive
black hole located at the GC. Taking a standard accretion efficiency
of 10\%, the black hole accretion rate can be determined $\dot{M}_{\rm
  BH}=2P_{\rm jet}/(0.1c^{2})=0.03$ $M_{\sun}/$yr (note that the black
hole ejects two opposing jets). During this AGN event ($0\leq t \leq
t_{\rm jet}=0.3$ Myr), the black hole accreted a total of $\Delta
{M}_{\rm BH}\sim 9000$ $M_{\sun}$ gas, and ejected in the jets a total
of $M_{\rm jet}\sim 1.5\times10^{5}$ $M_{\sun}$ gas, which is
estimated at $|z|=0.5$ kpc and thus includes the potential gas
entrained by the jets at $|z| <0.5$ kpc (see Table 2). Assuming that
Sgr A* has a mass of $M_{\rm BH}\sim 4\times
10^{6}$ $M_{\sun}$ \citep{ghez08}, the Eddington luminosity is $L_{\rm
  Edd}\sim 5.5\times 10^{44}$ erg s$^{-1}$, and the Eddington ratio
for the AGN activity is $\epsilon=2P_{\rm jet}/L_{\rm Edd}\sim
0.31$. However, as we discuss in Section 3.3, the energetics of the
\emph{Fermi} bubble event scales with the uncertain density of
the ambient halo gas, which confines the bubbles. For example, if the
normalization of the halo gas is lower by a factor of $10$, the energy
and power of the associated AGN event are also smaller by the same
factor $10$.

While the overall energetics, size, shape and age of our A1 \emph{Fermi}
bubble shown in the right panels of Figure 2 are encouraging, this
bubble is deficient in two important respects that can be expected
from any calculation based on the standard hydrodynamic equations 1 - 4.
First, the pagoda-shaped bubble boundaries visible in Figure 2 --
formed by a series of Kelvin-Helmholtz vortices -- disagrees
significantly with the globally smooth, egg-shaped gamma-ray bubbles
observed with the \emph{Fermi} telescope \citep{su10}.  Second, the
quasi-uniform distribution of CR electrons, proportional to $e_c(R,z)$
plotted in Figure 2, is inconsistent with the rather uniform gamma-ray
surface brightness observed by \emph{Fermi} across the bubble surfaces.  The
gamma-ray surface brightness produced by IC upscattering of rather
smoothly distributed ISRF photons is roughly proportional to the
integral of $e_c$ along the line of sight through the bubble.  For
spatially uniform $e_c$, the surface brighness is expected to decrease
significantly from the bubble center toward the boundaries at constant
$z$ where the ISRF photon density is roughly constant or decreases slightly in the same direction.  
Since the ISRF photon density is expected to decrease in the
$z$-direction (away from the Galactic center), $e_c(R,z)$ must also
increase in this direction.  Consequently, to produce a gamma-ray
image with uniform surface brightness, it is necessary that $e_c(R,z)$
increase smoothly toward all bubble boundaries. As we discuss below,
these two observational discrepancies suggest that additional physics plays a significant role
during the jet evolution. In Paper II, we show that these discrepancies can be corrected by 
including gas viscosity.

\subsection{Suppression of CR Diffusion across the Bubble Surface} 
\label{section:dsuppression}

\begin{figure*}
   \centering
  %  \epsscale{0.95}
\plottwo {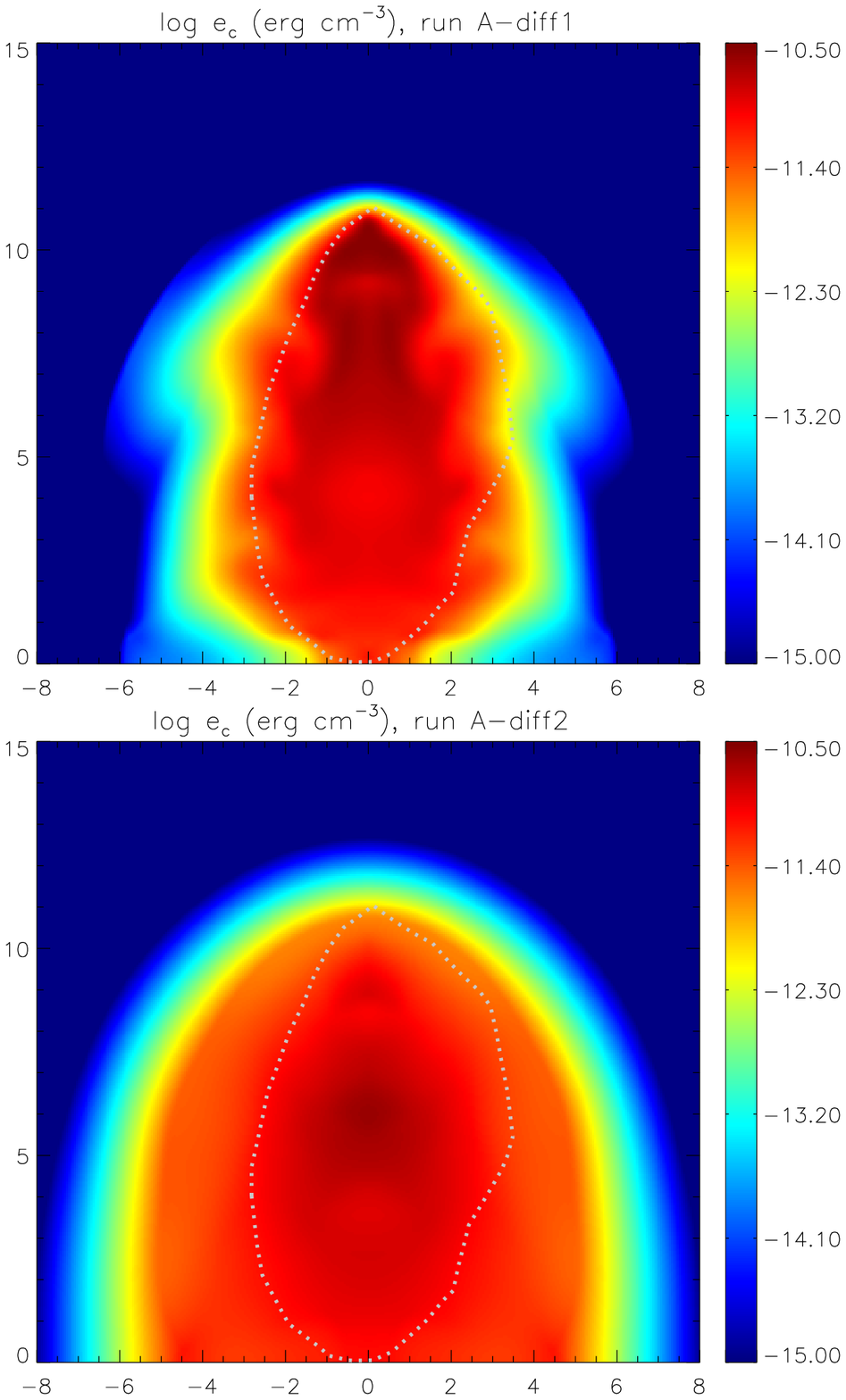} {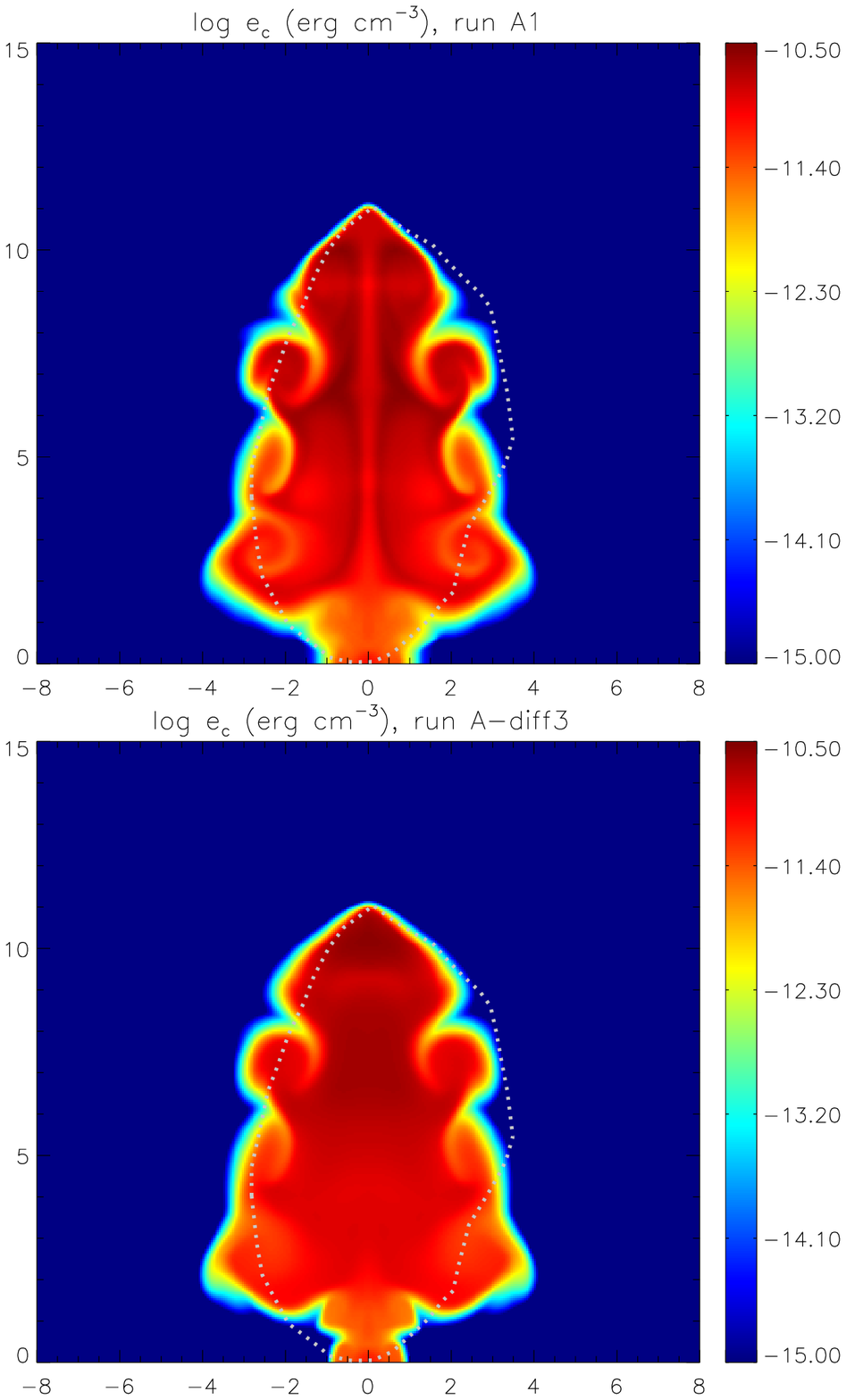} 
\caption{Central slices ($16\times15$ kpc) of CR energy density in
  logarithmic scale in run A-diff1 (top-left), A1 (top-right), A-diff2
  (bottom-left), and A-diff3 (bottom right) at $t=t_{\rm Fermi}$
  (details listed in Tables \ref{table1} and \ref{table2} for each
  run). Horizontal and vertical axes refer to $R$ and $z$
  respectively, labeled in kpc. The dotted region in each panel
  encloses the observed north \emph{Fermi} bubble. In runs A-diff1 and
  A-diff2, CR diffusion significantly affects the bubble evolution,
  rendering bubble edges that are less sharp than those
  observed. Variable CR diffusion in run A-diff3 leads to a smoother
  CR energy density distribution inside the bubble, while still
  suppressing CR diffusion across the bubble surface and retaining
  sharp bubble edges as in run A1.}
 \label{plot5}
 \end{figure*}  

 \begin{figure}
   \centering
 %   \epsscale{0.95}
\plotone {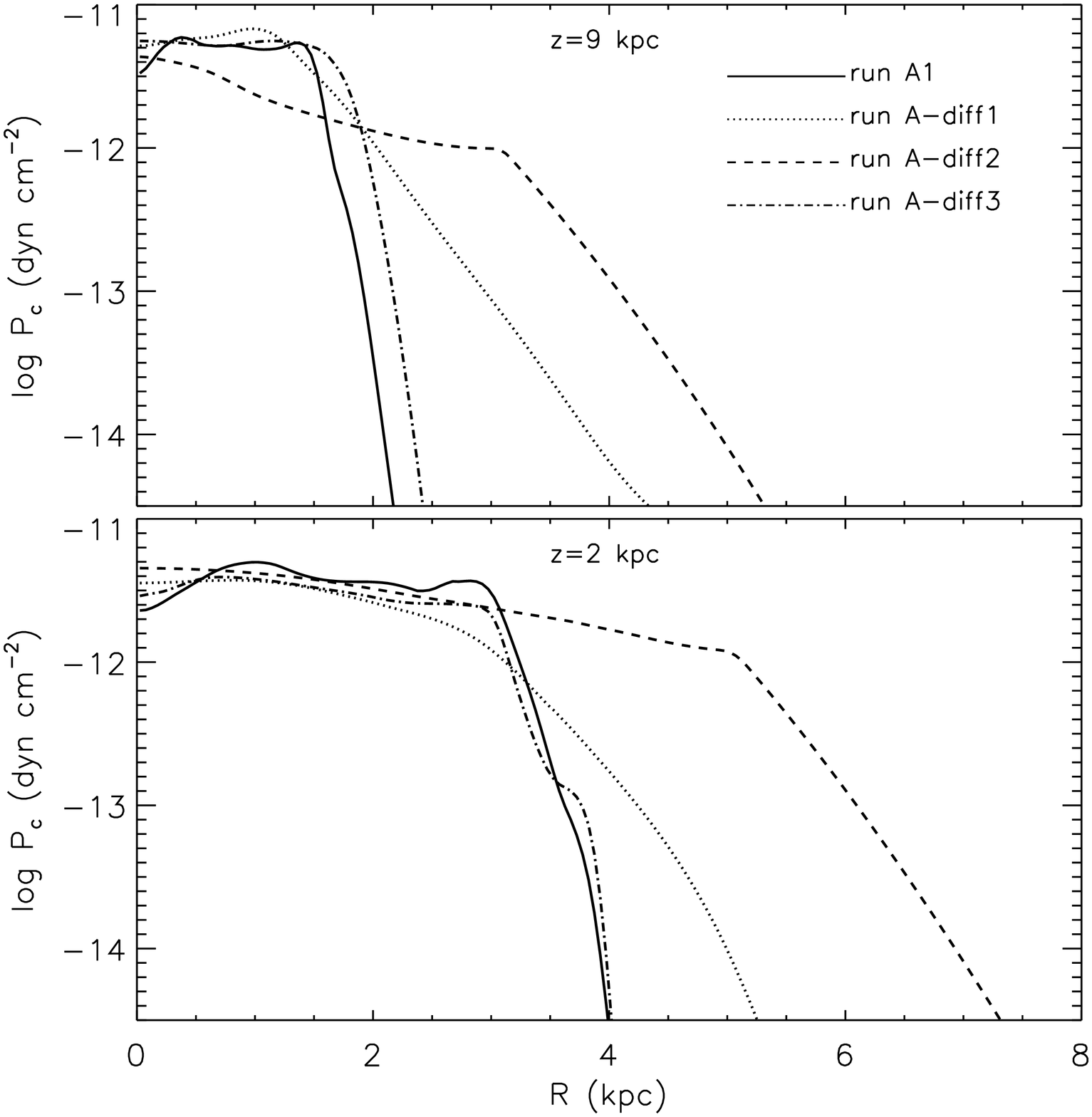} 
\caption{Variations of CR pressure $P_{\rm c}$ along the $R$-direction
  at $t=t_{\rm Fermi}$ at $z=2$ (bottom), and $9$ kpc (top) in runs A1
  (solid), A-diff1 (dotted), A-diff2 (dashed), and A-diff3
  (dot-dashed). The edge of the CR bubble is not sharp in runs A-diff1
  and A-diff2 because of CR diffusion across the bubble boundary.}
 \label{plot6}
 \end{figure} 
 
In our fiducial run A1, we choose a constant CR diffusion coefficient
$\kappa = 3\times 10^{27}$ cm$^{2}$ s$^{-1}$, which is much lower than
typical estimates of diffusivity in our Galaxy $\kappa \sim
(3-5)\times 10^{28}$ cm$^{2}$ s$^{-1}$. A low value for $\kappa$ is
specifically chosen to produce the CR bubble with sharp edge, which is
one of the main features in the observed \emph{Fermi} bubbles. When the
diffusivity is not strongly suppressed, the edges of the resulting CR
bubble are usually very broad, inconsistent with observations. To
study the effect of CR diffusion on the bubble evolution, we perform
three additional runs A-diff1, A-diff2, and A-diff3, which are all the
same as run A1, but with different prescriptions for the CR
diffusivity $\kappa$.

In run A-diff1 and A-diff2, the CR diffusion coefficient is chosen to
be $\kappa= 3\times 10^{28}$, and $3\times 10^{29}$cm$^{2}$ s$^{-1}$,
respectively. In these cases, diffusion helps transport CRs, and thus
the age of the CR bubble, $t_{\rm Fermi}=1.94$, $1.30$ Myr
respectively, is shorter than that in run A1. The 2D distribution of
CR energy density at $t=t_{\rm Fermi}$ in these runs is shown in left
panels of Figure \ref{plot5}. When compared with run A1 (right-top
panel), CR diffusion in runs A-diff1 and A-diff2 transports CRs to
much large distances in the $R$ direction, which is also clearly seen
in Figure \ref{plot6}, which shows variations of CR pressure $P_{\rm
  c}$ along the $R$-direction at $t=t_{\rm Fermi}$ at $z=2$ (bottom),
and $9$ kpc (top) in runs A1, A-diff1, A-diff2, and A-diff3.

An important feature of the CR bubble produced in runs A-diff1 and
A-diff2 is that the bubble edge is not sharp, as clearly seen in
Figures \ref{plot5} and \ref{plot6}. This feature is clearly produced
by CR diffusion, which transports CRs across the bubble surface,
rendering the {\it gradual} outward decline in CR energy density at
bubble edges. When the CR diffusion coefficient is significantly
suppressed as in our fiducial run A1, the CR bubble is mainly expanded
due to CR advection, and has sharp edges. Observations of \emph{Fermi}
bubbles show that the bubble edge is very sharp \citep{su10},
suggesting that the CR diffusion is significantly suppressed. However,
the CR diffusion coefficient may depend on the structure of local
magnetic fields and may vary with space and time. One interesting
question to ask is whether the sharpness of bubble edges implies that
the CR diffusion is suppressed everywhere in the bubble regions (e.g.,
as in run A1). It is possible that the CR diffusion inside the bubble
is not suppressed, as it does not directly affect the bubble
surface. The key to produce sharp bubble edges is the suppression of
CR diffusion across the bubble surface, which directly affects the
structure of bubble edges. Physically, due to the small gyro-radii of
relativistic particles (around $10^{-9}$ kpc for a $10$ GeV electron
in a $4$ $\mu$G field), CR diffusion may be anisotropic and CRs
diffuse mainly along magnetic field lines with cross-field diffusion
strongly suppressed. During the creation of the CR bubbles, 
ambient gas just external to the bubbles is 
strongly compressed, resulting in tangential magnetic
field lines near the surface. This may explain why CR diffusion is
suppressed across the bubble surface. It will be of great interest in
the future to study this issue in detail using magnetohydrodynamic
simulations with anisotropic CR diffusion.

Without explicitly introducing magnetic fields, here we want to do a preliminary
study investigating how the CR bubble evolves if the CR diffusion is
only suppressed across the bubble surface. To this end, we perform an
additional run A-diff3, where the CR diffusion is normal ($\kappa \sim
10^{28}$ - $10^{29}$ cm$^{2}$ s$^{-1}$) within the evolving CR bubble,
but significantly suppressed exterior to the bubble surface. As shown
in \S~\ref{section:formfermi}, the CR bubble is separated from the
surrounding thermal gas through a contact discontinuity, across which
thermal gas density changes abruptly. Thus we assume that in run
A-diff3, the CR diffusivity is related to thermal gas density through
an ad-hoc equation:
 \begin{eqnarray}
\kappa=
\begin{cases}
3\times 10^{29}(n_{{\rm e}0}/n_{\rm e}) \text{~cm}^{2} \text{~s}^{-1}    & ~ \text{when } n_{\rm e}> n_{{\rm e}0} \\
3\times 10^{29} \text{~cm}^{2} \text{~s}^{-1}    & ~ \text{when } n_{\rm e} \leq n_{{\rm e}0} \text{,}
\end{cases}
\label{equdiff}
\end{eqnarray}
\noindent
where $n_{{\rm e}0}=10^{-5}$ cm$^{-3}$. The parameters in equation
\ref{equdiff} are chosen so that during the calculation of this run
($t \leq t_{\rm Fermi}$), CR diffusion is always significantly
suppressed outside the expanding bubble ($\kappa \lesssim 10^{28}$
cm$^{2}$ s$^{-1}$), but not suppressed within it ($\kappa \sim
10^{28}$ - $10^{29}$ cm$^{2}$ s$^{-1}$). The low CR diffusivity
outside the bubble only suppresses CR diffusion across the bubble
surface and does not directly affect regions much further away from
the bubble, since there are essentially no CRs there. At $t=t_{\rm
  Fermi}$, the CR energy density distribution, shown in the
bottom-right panel of Figure \ref{plot5}, is very similar to that in
run A1 (top-right panel), and particularly, the edges of the CR bubble
are also very sharp, indicating that the prescription for CR
diffusivity shown in equation \ref{equdiff} indeed significantly
suppresses CR diffusion across the bubble surface, and CR diffusion in the bubble interior is not required to be suppressed to produce the sharpness of the bubble edges.
The bottom-right panel of
Figure \ref{plot5} also shows that the $e_{\rm c}$ distribution inside
the bubble in run A-diff3 is much smoother than in run A1. CR
diffusion inside the bubble removes local CR structures (e.g., regions
with high or low CR energy densities as seen in the right-top panel of
Fig. 5), which may otherwise have been seen in the \emph{Fermi} observations
of projected gamma-ray emission. The observed \emph{Fermi} bubbles have
approximately uniform surface brightness, which may imply that CR
diffusion is not strongly suppressed inside the bubbles (i.e., only
the CR diffusion across the bubble surface is strongly suppressed). Future data
from even longer-duration \emph{Fermi} observations are needed to study the
possible internal structure of the \emph{Fermi} bubbles.

\subsection{A Parameter Study and Degeneracies} 

\begin{figure*}
  \centering
  %  \epsscale{0.95}
\plottwo {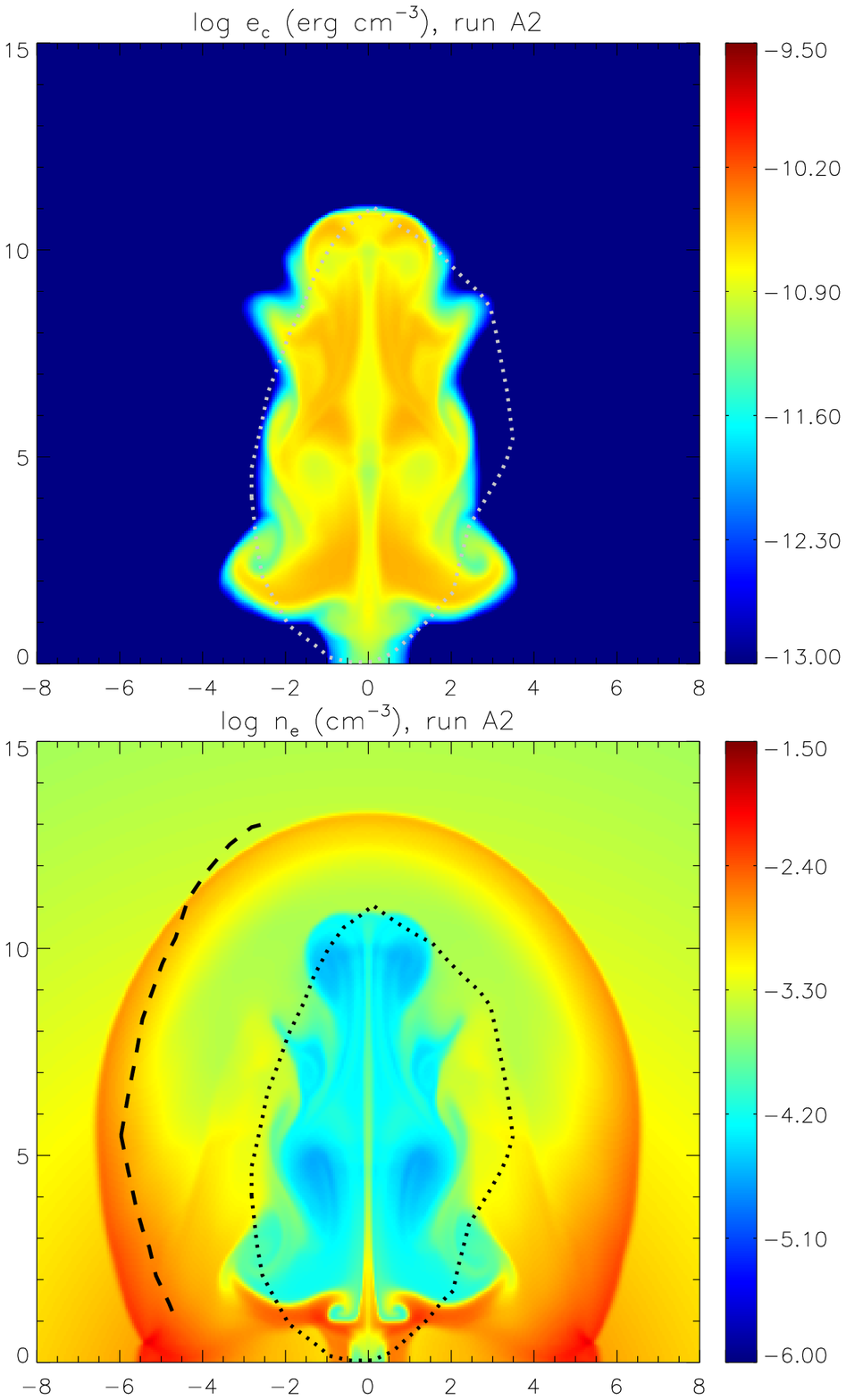} {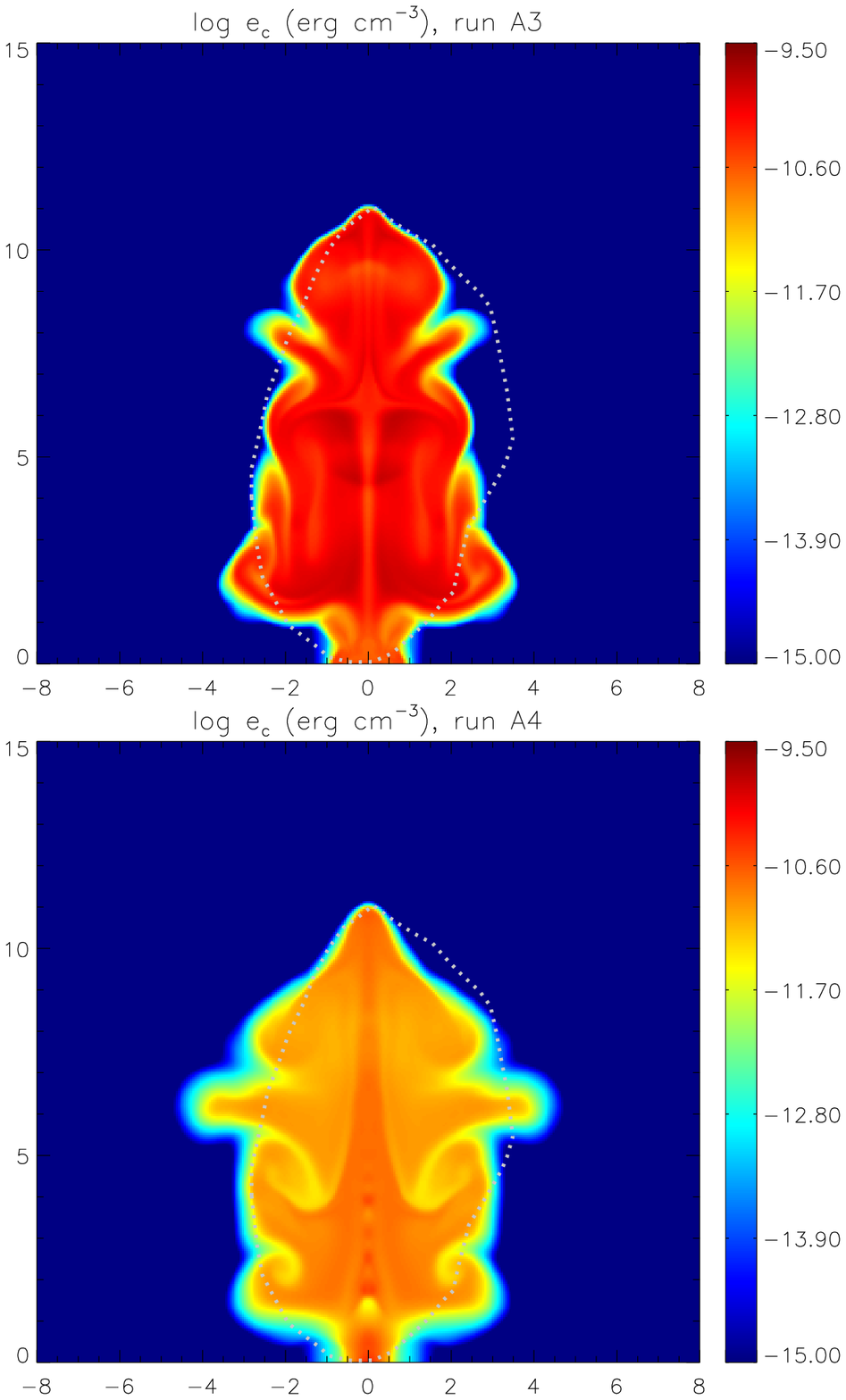} 
\caption{Central slices ($16\times15$ kpc) of log$(e_{\rm c})$ in run
  A2 (top-left), A3 (top-right), A4 (bottom-right), and log$(n_{\rm
    e})$ in run A2 (bottom left) at $t=t_{\rm Fermi}$ (details listed
  in Tables \ref{table1} and \ref{table2} for each run). Horizontal
  and vertical axes refer to $R$ and $z$ respectively, labeled in
  kpc. The dotted region in each panel encloses the observed north
  \emph{Fermi} bubble. The dashed lines in the bottom-left panel trace the
  outer edge of the {\it ROSAT} X-ray emission feature surrounding the
  northern bubble. Run A1 and these additional runs have different jet
  parameters, but all produce CR bubbles approximately matching the
  observed \emph{Fermi} bubbles.}
 \label{plot7}
 \end{figure*}  
 
 \begin{figure}
   \centering
  %  \epsscale{0.5}
\plotone {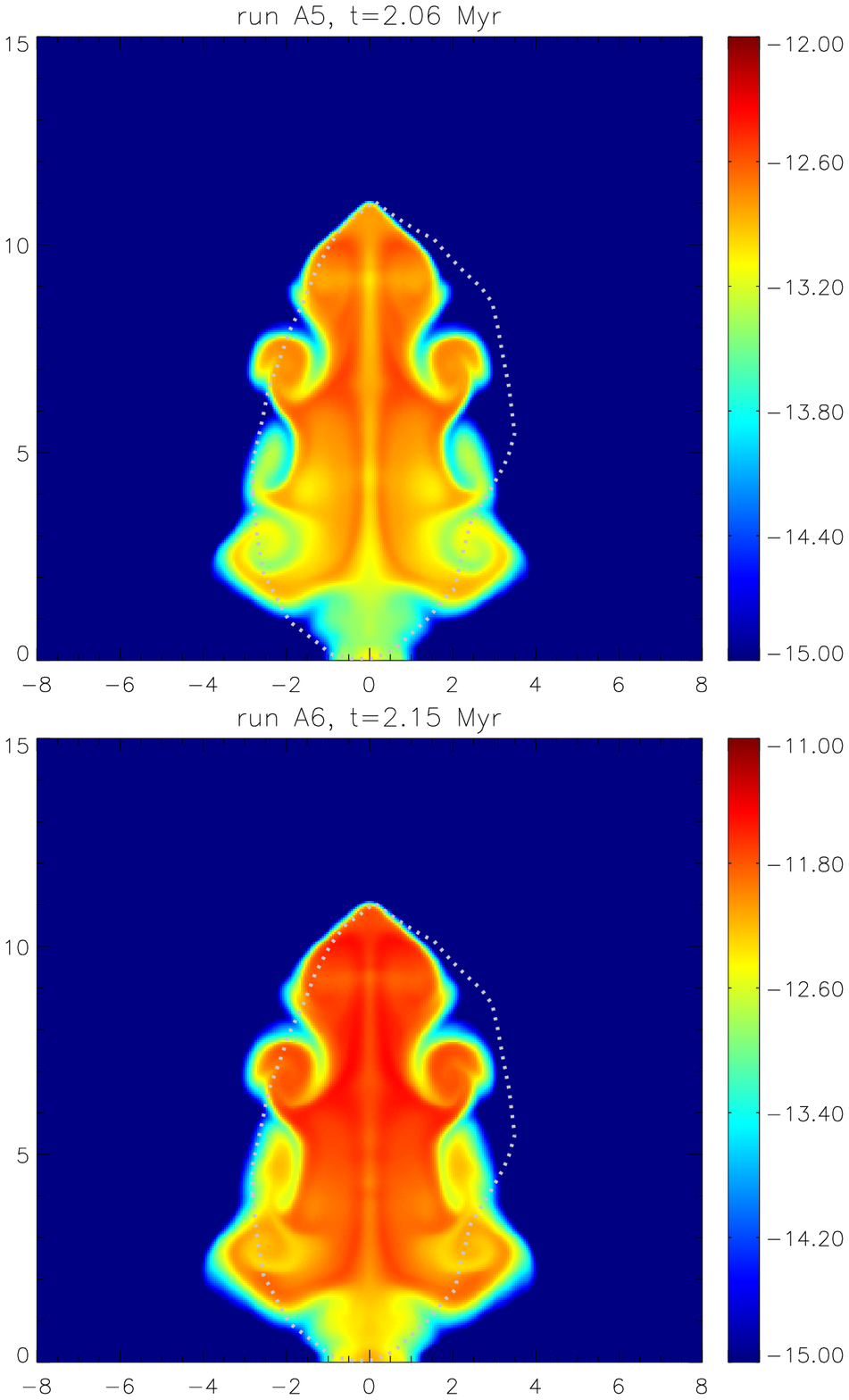} 
\caption{Central slices ($16\times15$ kpc) of log$(e_{\rm c})$ in run
  A5 (top) and A6 (bottom) at $t=t_{\rm Fermi}$. Horizontal and
  vertical axes refer to $R$ and $z$ respectively, labeled in kpc. The
  dotted region in each panel encloses the observed north \emph{Fermi}
  bubble. Run A5 shows that the required jet power and the CR energy
  density are proportionally smaller in a gaseous halo with lower
  densities compared to run A1, while run A6 shows that the
  morphological bubble evolution is quite insensitive to the relative
  contribution of CRs and thermal gas to the jet pressure when the
  total jet pressure and other model parameters are fixed.  }
 \label{plot8}
 \end{figure} 
 
The jet simulation involves a large number of parameters, many of
which are degenerate. It is thus impossible to uniquely determine all
jet properties from the morphology of the observed \emph{Fermi} bubbles. For
example, a shorter jet duration $t_{\rm jet}$ results in a longer
bubble age and a `fatter' bubble, but this effect can be averted by a
larger jet density $n_{\rm ej}$, a higher jet velocity $v_{\rm jet}$,
or a smaller CR energy density $e_{\rm jcr}$ in the jet. Here we
present results of a few additional (essentially degenerate) runs (A2,
A3, and A4), all of which roughly produce the morphology of \emph{Fermi}
bubbles. The jet parameters and some results are summarized in Tables
\ref{table1} and \ref{table2}. Compared to run A1, the jet is more
massive in run A2 ($\eta=0.02$), faster in run A3 ($v_{\rm
  jet}=0.2c$), and narrower in run A4 ($R_{\rm jet}=0.2$ kpc). The age
of \emph{Fermi} bubbles in these runs is $t_{\rm Fermi}=1.74$, $0.86$, $2.34$
Myr respectively. The 2D distribution of CR energy density at
$t=t_{\rm Fermi}$ in these runs is shown in Figure \ref{plot7}, where
the electron number density distribution in run A2 is also shown in
the bottom-left panel. In run A2, the shock fits even better with the
outer edge of the X-ray shell feature (dashed line) than in run A1,
strengthening the point that the {\it ROSAT} X-ray features
surrounding the \emph{Fermi} bubbles are produced by shocked gas associated
with the jet activity. These four runs are representative of
successful jet models which roughly reproduce the overall size and shape of the \emph{Fermi} bubbles, suggesting that the \emph{Fermi} bubbles could be
produced by a powerful AGN jet event which began about $1$ - $3$ Myr
ago, and was active for a duration of $\sim 0.1$ - $0.5$ Myr.

The evolution of AGN jets and the resulting CR bubbles also depend on
the initial density of the confining halo gas. For a gaseous halo with
fixed temperature, its initial hydrostatic density distribution can be
scaled up or down by varying the central gas density $n_{\rm
  e0}$. Equations \ref{hydro1} - \ref{hydro4} ensure that the
evolution of the jet and resulting CR bubble is the same if the
following three jet properties -- density $\rho_{\rm j}$, thermal
energy density $e_{\rm j}$ and CR energy density $e_{ \rm jcr}$ are
scaled by the same factor as $n_{\rm e0}$. As an example, here we
present the result of an additional run A5, where $n_{\rm e0}$,
$\rho_{\rm j}$, $e_{\rm j}$, and $e_{ \rm jcr}$ are all scaled down by
the same factor $100$ with respect to run A1. The spatial distribution
of CR energy density in this run at time $t=t_{\rm Fermi}$ is shown in
the top panel of Figure \ref{plot8}, confirming that the morphological
evolution of the jet and bubble is the same as in run A1. It is
important to note that the jet duration and the current age of the
\emph{Fermi} bubbles are not affected by the halo gas density, while the
total jet energy and the CR energy density within the bubbles scale
with it. Thus, due to our limited knowledge of the halo gas density,
the energetics of the AGN event responsible for the \emph{Fermi} bubbles --
as well as the CR energy density and the gamma-ray emissivity -- is
not constrained very well by our model. Depending on the initial density distribution 
in the Galactic halo, the energetics of the  \emph{Fermi} bubble event 
may be $\sim 10^{55}$ - $10^{57}$ erg (Table 2).

In principle, accurate X-ray observations of the shell of shocked gas
surrounding the \emph{Fermi} bubbles can improve our estimate of the gas
density distribution in the pre-bubble atmosphere and consequently the
energetics of the dynamical event that created the bubbles.  All-sky
{\it ROSAT} X-ray observations at 0.75 and 1.5 keV were used by
\citet{sofue00} to identify dumbbell-shaped shells of soft X-ray emission
having lobes about 10 kpc in diameter symmetric about the Galactic
center and aligned approximately along the Galactic rotation axis.  At
the time \citet{sofue00} interpreted this as emission from gas in the
Galactic halo that was shocked by an explosion in the Galactic center
with total energy $E_{\rm tot} \sim 10^{56}$ erg that occurred about $1.5
\times 10^7$ yr ago.  However the original atmosphere adopted by
\citet{sofue00} is considerably different (not in hydrostatic equilibrium) and about 18 times less dense
than ours at $R = 0$, $z = 5$ kpc (Figure \ref{plot1}) near the bubble
center. The corresponding single-bubble energy found by
\citet{sofue00}, when scaled to our atmosphere, is $E_{\rm jet} =
0.5E_{\rm tot} \times 18 \approx 9 \times 10^{56}$ erg, which is consistent
with our model A1 where $E_{\rm jet} = 8 \times 10^{56}$ erg.  
Similar to the \emph{Fermi} bubbles, the real energetics of the {\it ROSAT} X-ray shells
is not well constrained due to the uncertain density of the original gas in the Galactic halo.
However, if they were indeed produced by a jet event (e.g. the \emph{Fermi} bubble event),
their age may be much shorter than 15 Myrs as estimated in a starburst model by \citet{sofue00}.

The evolution of AGN jets and resulting bubbles depends on the total pressure 
in the initial jet, but it is insensitive to how the jet pressure is distributed between 
thermal and non-thermal components. 
Thus, another uncertainty in our dynamical model is the ratio of the CR to thermal
pressure within the jet, which is directly related to an important
problem -- the particle content (thermal versus non-thermal) within the current
\emph{Fermi} bubbles. In run A1, the CR pressure dominates in the initial AGN
jet (the ratio of CR to thermal pressure is $9.25$), resulting in
comparable CR and thermal pressures within the \emph{Fermi} bubbles at the current time (see the
bottom panel in Fig. \ref{plot4}). However, the CR-to-thermal pressure ratio in the bubbles depends on 
its initial value at the jet base, and is generally not $1$ in our simulations.
In an additional run A6, we follow
the evolution of a jet dominated by thermal pressure (the initial ratio of CR
to thermal pressure is $\sim 0.1$; the total jet pressure, jet density
and other parameters are the same as run A1; see Table \ref{table1}),
finding that the morphological evolution of the resulting CR bubbles
in this run is very similar to run A1, as clearly seen in the bottom
panel of Figure \ref{plot8}. Due to much lower CR power in the initial jet, 
run A6 produces \emph{Fermi} bubbles having proportionally
lower CR pressure $P_{\rm c}$ (much lower than thermal pressure) and lower gamma ray surface
brightness than our fiducial run A1. The CR pressure in our successful runs 
by no means represents the real CR pressure in the \emph{Fermi} bubbles, which 
can be constrained by microwave and gamma ray emissions from the  \emph{Fermi} bubbles (see Sec. 3.5).
However, our dynamical model is robust and the bubble evolution, which depends on the 
total jet pressure, is insensitive to the CR pressure. Analyses based on
multi-wavelength (X-ray, gamma-ray, microwave, radio) observations of
the \emph{Fermi} bubbles and the surrounding gas, which is beyond the scope
of this paper, should be used to constrain the density of halo gas and
the particle content of the \emph{Fermi} bubbles.

\subsection{Constraints on the Jet Density} 
 
\begin{figure}
  \centering
%    \epsscale{0.5}
\plotone {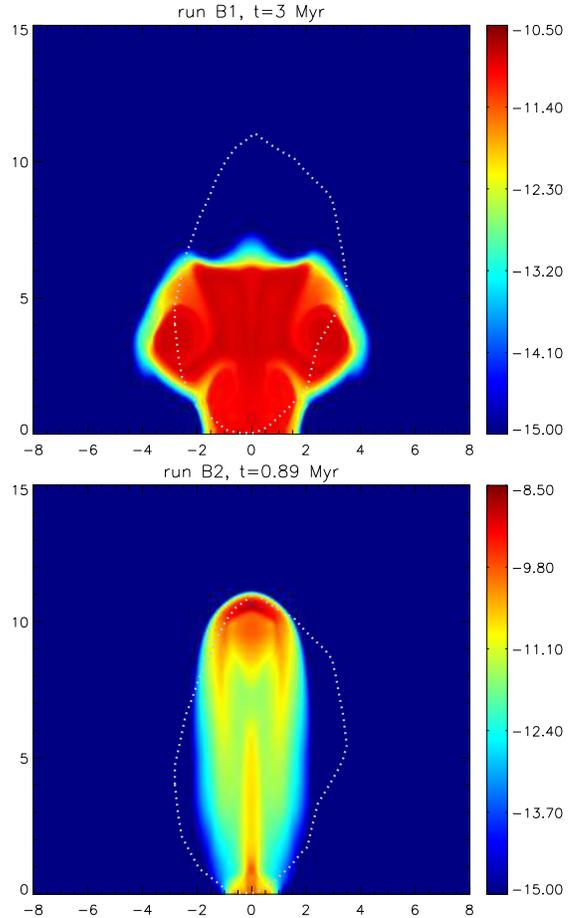} 
\caption{Central slices ($16\times15$ kpc) of CR energy density in
  logarithmic scale in run B1 at $t=3$ Myr (top) and B2 (bottom) at
  $t=t_{\rm Fermi}$. Horizontal and vertical axes refer to $R$ and $z$
  respectively, labeled in kpc.The dotted region in each panel
  encloses the observed north \emph{Fermi} bubble. Very light jets in run B1
  ($\eta=0.0001$) decelerate rapidly in the halo gas, forming
  quasi-spherical CR bubbles unlike the radially elongated \emph{Fermi}
  bubbles observed. On the other hand, the massive jets in run B2
  ($\eta=0.5$) transport most CRs just along the $z$ axis, not forming
  the 'fat' bubbles observed.}
 \label{plot9}
 \end{figure}

As discussed in the previous subsection, it is hard to accurately
determine the jet properties, but it is possible to put some useful
constraints, particularly on the jet density. When the jet is very
light, e.g., $\eta=0.0001$ as in run B1 (see Table \ref{table1} for
other jet parameters), it quickly decelerates in the hot gas, and
expands laterally in the $R$ direction due to the high pressure within
the bubble. The top panel of Figure \ref{plot9} shows the CR energy
density in logarithmic scale at $t=3$ Myr. As clearly seen, the jet
only reaches $z\sim 6$ kpc at this time, and the resulting CR bubble
is quasi-spherical, unlike the observed \emph{Fermi} bubbles elongated in the
$z$ direction. Very light jets tend to have strong backflows (i.e.,
with $v_{\rm z} < 0$ inside the bubble at large $R$), which expand
laterally, forming CR bubbles much `fatter' than \emph{Fermi} bubbles (see
\citealt{guo11}). We have experimented with more runs with different
jet densities, and find that the \emph{Fermi} bubbles can be successfully
reproduced only with jet densities $\eta \gtrsim 0.001$.

On the other hand, when the jet is very massive, e.g., $\eta=0.5$ as
in run B2 (see the bottom panel of Figure \ref{plot9}), it decelerates
very slowly, producing very little backflow. In this case, most of the
CRs are advected by the jet to the jet tip, not forming a `fat',
radially-elongated bubble as observed. We have run many simulations
with high jet densities, and found that the formed CR bubbles are usually
much thinner than the observed \emph{Fermi} bubbles. Acceptable runs
typically have the jet density contrast $\eta \lesssim 0.1$.

AGN jets efficiently transport CRs from the GC to the Galactic
halo. Additionally the jets can produce or reaccelerate CRs in the
strong shock at the jet tip (hot spot) which we have not considered
here. The morphology and evolution of the resulting CR bubble
significantly depend on many jet properties, but the jet density
contrast is particularly important. From the morphology of the \emph{Fermi}
bubbles, we can approximately constrain the jet density contrast:
$0.001 \lesssim \eta \lesssim 0.1$, which corresponds to $6 \times
10^{-5} \text{ cm}^{-3} \lesssim n_{\rm ej} \lesssim 6 \times
10^{-3}\text{ cm}^{-3}$ for our adopted model for the Galactic thermal
atmosphere. The constraint on the jet density contrast $\eta$ is quite
robust, not depending on the value of $n_{\rm e0}$, the density
normalization of the ambient halo gas.

\subsection{Success and Problems of Our Jet Model} 

The detection of the \emph{Fermi} bubbles in the Galaxy is one of the biggest
discoveries in astronomy in recent years. The origin of the bubbles
remains mysterious. Here we are proposing that the bubbles were
produced by a recent powerful AGN jet event from the GC. We focus on
the formation and dynamical evolution of the \emph{Fermi} bubbles, leaving
details of the multi-wavelength emission features to future work. In this
subsection, we summarize the success and potential problems of our jet
model in explaining the observed features of the \emph{Fermi} bubbles.

Our model successfully explains a few key observational features of
the \emph{Fermi} bubbles. (1) The origin of CRs is naturally explained within a
jet scenario. CRs can be accelerated to very high energies in AGN jets
near supermassive black holes and/or in jet hotspots, as seen in
numerous observations of extragalactic AGN jets. (2) Sub-relativistic jets naturally produce large
CR bubbles within a few Myrs, {\it a short bubble age} inferred from
the fact that the gamma-ray emission is likely dominated by the IC
emission of CRes (\citealt{dobler10} and \citealt{su10}). It
is not trivial to transport or accelerate CRs in 10 kpc sized bubbles
within such a short amount of time. (2) The opposing jet scenario
naturally explains the bilobular morphology of the bubbles, which are
symmetric with respect to the GC. (3) We identified a few sets of jet
parameters, with which the jets produce CR bubbles with morphologies similar
to the observed bubbles (mainly the bubble elongation and
ellipticity). (5) In our model, the jet duration is much shorter than the bubble age, implying that the
CRs in the whole bubble have similar age at the current time. This
explains the {\it Fermi} observation that the gamma ray spectral index
is quite uniform across most regions of the two bubbles. (6) The jet
event produces a strong shock, which heats and compresses the ambient
gas in the Galactic halo, potentially explaining the {\it ROSAT}
X-ray shell features surrounding the bubbles. (7) Sharp bubble edges indicate that the \emph{Fermi} 
bubbles have been expanding mainly due to CR advection, while CR diffusion across bubble edges
is strongly suppressed.

\begin{figure}
  % \epsscale{0.5}
\plotone {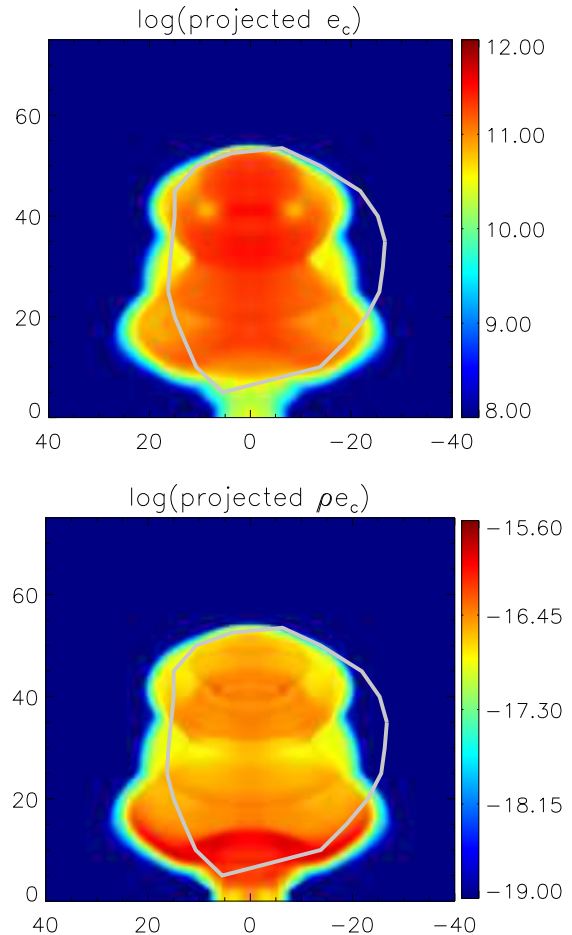} 
\caption{Line-of-sight projection of CR energy density (top; a proxy for the gamma-ray surface brightness due to CR electrons) and $\rho e_{\rm c}$ (bottom; a proxy for the gamma-ray surface brightness originated from CR protons) in logarithmic scale in run A1 at $t=2.06$ Myr. Horizontal and vertical axes refer to Galactic longitude and latitude respectively, labeled in degrees. The solid circle represents the edge of the observed south \emph{Fermi} bubble. Edge irregularities and edge darkening are clearly seen at high latitudes. At low latitudes ($|b|\lesssim 20^{\circ} - 30^{\circ}$), \emph{Fermi} observations are significantly contaminated by emissions from the Galactic disk and bulge. The spatial variations of CR spectra and ISRF may affect the gamma-ray map, which is not considered here.}
 \label{plot10}
 \end{figure}

However, there are still some potential problems associated with our
model, which need to be investigated in future studies. The main
difference between the simulated CR bubble and the observed \emph{Fermi}
bubbles is that the former suffers from Kelvin-Helmholtz instabilities at its surface (as clearly seen in
Figure \ref{plot2}), while the real \emph{Fermi} bubbles have smooth
edges. Line-of-sight projection tends to smooth the bubble edge when
viewed in Galactic coordinates, but it is unable to completely remove
surface irregularities created by these instabilities, as seen in Figure \ref{plot10}, 
which shows for run A1 the line-of-sight projection of CR energy density 
(a proxy for the gamma-ray surface brightness due to CR electrons) and $\rho e_{\rm c}$ (a proxy for the gamma-ray surface brightness originated from CR protons). This
discrepancy suggests that there may be additional physics, which plays
an important role in the bubble evolution, suppressing the development
of these instabilities. Similar surface instabilities have received much
attention in numerical studies of radio bubbles and X-ray cavities in
galaxy clusters, where gas viscosity \citep{reynolds05} and magnetic
tension \citep{ruszkowski07}, have been invoked to successfully
suppress these instabilities at the boundaries of buoyantly-rising
bubbles. But it remains to be shown that the instabilities can also
be suppressed during the creation of AGN bubbles. 
A preliminary study exploring the beneficial role
of viscosity on the bubble evolution and morphology is presented in Paper II.
Viscous suppression of instabilities in the \emph{Fermi} bubbles may provide an
unusual opportunity to study the interesting gas microphysics -- viscosity -- in hot
plasma.

An interesting feature of the observed \emph{Fermi} bubbles is the
approximately uniform gamma-ray surface brightness, particularly at
high latitudes ($|b|\gtrsim 30^{\circ}$).  However, a nearly uniform
CR distribution, as in our models computed here, produces a
center-brightened (limb-darkened) surface brightness distribution 
in projection (see Figure \ref{plot10}), which
is inconsistent with uniform gamma ray brightness observed in the
\emph{Fermi} bubbles. A flat surface brightness distribution implies that the
CR density increases gradually toward bubble edges and with Galactic
latitude from the bubble center. However, if CRs are too concentrated
at the bubble surface, the gamma-ray surface brightness will be too
strongly limb-brightened. Thus, the CR distribution may need some
fine-tuning to get a roughly flat gamma-ray intensity.  Although such
an edge-favored CR distribution is not reproduced in our current jet
simulations, this discrepancy does not invalidate the jet scenario for
the \emph{Fermi} bubbles. Instead, it may require a new physical
mechanism that operates additionally in the standard AGN jet model.  
In Paper II, we argue that hot gas viscosity may provide the additional physics to suppress
surface irregularities and to produce an edge-favored CR distribution, 
resulting in a more uniform gamma ray
surface brightness distribution.  Other physical mechanisms (e.g.,
stochastic re-acceleration of CR electrons preferentially near bubble
edges as suggested by \citealt{mertsch11}) may also contribute to
surface brightness uniformity.

In this paper, we choose a very simple model for the ambient gas distribution --
a hot volume-filling isothermal gas in hydrostatic equilibrium. In reality, the gas
distribution in the Galaxy may be much more complex. 
Gas motions (e.g. winds) may be present, producing non-axisymetric features in the bubble morphology.
At low latitude, the hot, warm and cold gas
within and near the Galactic bulge may significantly affect the jet evolution and the bubble
shape there. Some of our current simulations produce a `stem' feature at low latitude (see Fig. \ref{plot2}),
which can not be accurately studied without modeling the multi-phase gas there. 
Due to strong gamma-ray contaminations from the Galactic disk and bulge at low latitude,
further studies are required to investigate if the stem is indeed present or not in the observed \emph{Fermi} bubbles.

 \begin{figure}
% \epsscale{0.9}
\plotone {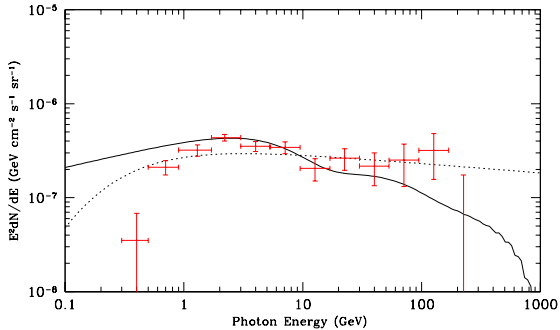} 
\caption{Gamma-ray spectrum from a CR electron population (IC scattering; solid line) and a CR proton population (pion decay; dashed line). The CRe spectrum is taken to be a power law in energy: $dN/dE \propto E^{-2.4}$ with $0.1 \leq E \leq 1000$ GeV. The CR proton spectrum is assumed to be a power-law in momentum: $dN/dp \propto p^{-2.1}$ with $0.1 \leq p/m_{\rm p}c \leq 1000$. Crosses represent the observed average gamma ray flux from the \emph{Fermi} bubbles \citep{su10}.}
 \label{plot11}
 \end{figure} 
 
With hydrodynamic simulations, we focus on the evolution and morphology of the \emph{Fermi} bubbles in this paper. Our simulations also investigate the transport of CRs (advection and diffusion) by following the evolution of the CR energy density $e_{\rm c}$, which may be contributed by CR electrons and/or protons with arbitrary spectra. The real CR content and spectrum in the \emph{Fermi} bubbles may be constrained by the observed gamma ray spectrum, which is being actively studied and debated in the literature (e.g. \citealt{su10}; \citealt{crocker11}). Here we present simple emission calculations in two extreme cases, assuming that the gamma ray flux is emitted entirely by CR electrons and protons respectively. In the leptonic scenario, we follow \citet{su10}, adopting a power-law electron spectrum $dN/dE \propto E^{-2.4}$ with $0.1 \leq E \leq 1000$ GeV. We choose the normalization of the electron spectrum so that the calculated gamma ray spectrum, which is shown as the solid line in Figure \ref{plot11}, is roughly consistent with the observed gamma ray luminosity. Here we used the Klein-Nishina IC cross section \citep{blumenthal70} and an approximate line-of-sight integration length of $4$ kpc across the bubbles. The ISRF model is taken from the latest version (version 54.1.984) of GALPROP (evaluated at the bubble center $R=0$, $|z|=5$ kpc). We also calculated synchrotron emission from CR electrons in this model and found that, when the magnetic field strength in the bubbles is chosen to be $8$ $\mu$G, the synchrotron flux at $23$ GHz is around $1.1$ kJy/sr, consistent with the detected WMAP flux at around $b\sim -20^{\circ}$ - $-30^{\circ}$. It remains unclear why the microwave flux from the south bubble drops significantly at $b\lesssim -35^{\circ}$ (see \citealt{dobler12}). 

Combining with our simulations, emission calculations may provide new insights on the \emph{Fermi} bubbles. To this end, we calculated the CR pressure contributed by electrons between $0.1$ and $1000$ GeV in the above leptonic model and found that it is around $P_{\rm c}=2.0 \times 10^{-15}$ dyn cm$^{-2}$, which is about four orders of magnitude below the total pressure ($\sim 10^{-11}$ dyn cm$^{-2}$) inside the simulated \emph{Fermi} bubbles in run A1 (see Figs \ref{plot3} and \ref{plot4}). This result is quite robust with respect to the power index of the electron spectrum, suggesting that the pressure within the \emph{Fermi} bubbles is not dominated by CR electrons in the leptonic scenario. The dominant bubble pressure may instead come from thermal gas (e.g. in run A6), CR protons, or magnetic fields.

In the hadronic scenario, the gamma ray emission of the \emph{Fermi} bubbles is dominated by the decay of neutral pions produced from hadronic collisions of CR protons with thermal nuclei (\citealt{crocker11}). We assume that CR protons have a power-law distribution in momentum: $dN/dp \propto p^{-2.1}$ with $0.1 \leq p/m_{\rm p}c \leq 1000$. We use an analytic formula in \citet[eq. 62]{ensslin07} to approximate gamma-ray emissivity from pion-decay and adopt thermal electron density in the \emph{Fermi} bubbles to be $10^{-4}$ cm$^{-3}$ (as in run A1; Figs \ref{plot3} and \ref{plot4}), which was used to derive the target nucleon density. The resulting gamma-ray flux is shown in Figure \ref{plot11} (dotted line) and the CR proton pressure in this model is $P_{\rm c}=5.1 \times 10^{-11}$ dyn cm$^{-2}$, slightly larger than the total pressure ($\sim 10^{-11}$ dyn cm$^{-2}$) inside the simulated \emph{Fermi} bubbles in run A1. This does not rule out the hadronic scenario, since (1) the required CR proton pressure drops if the thermal gas density in the bubbles is higher and (2) the total bubble pressure in our dynamical simulations increases if the initial ambient gas pressure is larger. However, our simple calculations suggest that the required CR proton pressure in the hadronic scenario may dominate the total pressure in the bubbles, and is much larger than the CR electron pressure in the leptonic scenario.

 \section{Summary and Implication}
\label{section:conclusion}

The detection of \emph{Fermi} bubbles in the Galaxy is one of the most
important and striking findings during the first two years'
observations of the {\it Fermi Gamma-ray Space Telescope}. The bubbles
are symmetric about the GC, with one above and the other below the
Galactic plane. The surface brightness in $\gamma$-ray emission from
the bubbles is quite uniform with sharp edges at the bubble
boundaries. These observed properties make it very difficult to
explain the bubbles in many of the standard ways, for example by
supernova shocks in the Galactic plane or by dark matter
annihilations.

The unique location, morphology, and sharp gamma-ray edges of the bubbles suggest that they
were probably created by an episode of energy injection in
the GC. Gamma rays from the bubbles can be produced by CR electrons and/or protons. 
Motivated by numerous double radio sources and
X-ray cavities observed in other galaxies, we investigate in this paper if AGN jets from the GC
can create the \emph{Fermi} bubbles with the observed 
morphology, size and appropriate age.
In our model, a pair of bipolar jets were released from the GC along
the rotation axis of the Galaxy. We model the
jet evolution using a series of 2D axisymmetric hydrodynamical
simulations, following the evolution of CRs and their
dynamical interactions with thermal gas (see Section
\ref{section:equation}). CR advection is modeled 
self-consistently and CR diffusion is also considered.

We show that the observed \emph{Fermi} bubbles could be reproduced by a recent
AGN jet event about $1$ - $3$ Myr ago, which was active for a
duration of $\sim 0.1$ - $0.5$ Myr. The total jet energy released
during the AGN event may be $\sim 10^{55}$ - $10^{57}$ erg,
depending on the initial density distribution 
of the hot halo gas which confines the bubbles. Containing both thermal gas
and CRs, the two-fluid jet quickly advects CRs to high latitude. The
jet is energetically dominated by the kinetic energy, and
over-pressured with either CR or thermal pressure which induces
lateral expansion, creating a fat CR bubble as observed. In our
fiducial run A1, the two opposing jets have a total power of $2P_{\rm
  jet}\sim 1.7 \times 10^{44}$ erg s$^{-1}$, corresponding to an
accretion rate of $\dot{M}_{\rm BH} \sim 0.03$ $M_{\sun}/$yr for the
central black hole and an Eddington ratio of $\epsilon \sim 0.31$. The
jet activity also induces a strong dumbbell-shaped shock (currently Mach
number $M\sim 8$ - $9$), which heats and compresses the ambient gas in
the Galactic halo, qualitatively similar to the dumbbell-shaped X-ray
shell features observed by the {\it ROSAT} X-ray telescope.

The two observed \emph{Fermi} bubbles are roughly located and
elongated along the rotation axis of the Galaxy.
In the jet scenario, this morphology requires that the two opposing jets that produced
the \emph{Fermi} bubbles were ejected nearly along the Galactic rotation axis, as assumed in our simulations.
This jet direction may be possible if the gas accreted by Sgr A* had angular momentum in this
direction or Sgr A* had a relatively strong spin along this direction before the accretion.
It is unclear if this special orientation observed in the \emph{Fermi} bubbles is a coincidence or a general feature of CR bubbles in disk galaxies (see \citealt{baum93} for the same preferential orientation detected in radio bubbles in Seyfert galaxies).

Because of the degeneracy of successful bubble models with different
jet parameters, it is difficult or impossible to determine unique jet
parameters directly from the bubble location and morphology. However,
we can put some useful constraints on them, particularly on the jet
density contrast relative to the ambient hot gas at the jet base. Very
low density, ultra-light jets decelerate quickly and usually form
quasi-spherical bubbles unlike the observed \emph{Fermi} bubbles which are
radially elongated. In contrast, heavy, massive jets decelerate very
slowly, advecting most CRs to the tip of the jet with too little
lateral expansion. Successful jets are moderately light, having
typical density contrasts $0.001 \lesssim \eta \lesssim 0.1$ (i.e., $6
\times 10^{-5} \text{ cm}^{-3} \lesssim n_{\rm ej} \lesssim 6 \times
10^{-3}\text{ cm}^{-3}$ for our adopted model for the Galactic thermal
atmosphere).

We also show that to produce sharp edges of the bubbles, CR diffusion
across the bubble surface must be suppressed 
significantly below the CR diffusion rate observed in 
the solar vicinity. CR advection is
responsible for the spatial expansion of the \emph{Fermi} bubbles, which
compresses thermal gas near the bubble surface, aligning the local
magnetic fields to be tangential to the bubble boundary. CR diffusion
is probably anisotropic, with cross-field diffusion strongly
suppressed. However, it is likely that CR diffusion deeper inside the
bubbles is not suppressed and it tends to remove small CR structures,
resulting in a smoother CR distribution and $\gamma$-ray emission
within the bubbles as suggested by \emph{Fermi} observations.

The total pressure inside the \emph{Fermi} Bubbles must not be less than the pressure of CRs required to produce the gamma ray emission observed. If the gamma-ray emission from the bubbles is mainly due to CR electrons, the required CR electron pressure to produce the observed gamma-ray flux is negligible compared to the total bubble pressure, which may instead be dominated by other components, e.g., thermal gas, CR protons, or magnetic fields. On the other hand, if the gamma-ray emission is mainly due to CR protons, the required CR proton pressure is much higher, probably dominating the total bubble pressure.

The \emph{Fermi} bubbles provide plausible evidence for a recent powerful AGN jet
activity in the Milky Way. Extragalactic jets have been detected in
many distant galaxies, such as the famous jet in M87
(\citealt{junor99}; \citealt{kovalev07}). Numerous bipolar radio
bubbles and X-ray cavities have also been observed, clearly associated
with the central stellar bulges of massive galaxies. These non-thermal
regions are almost certainly produced by AGN jets, which are often not
detectable because of the short time when they are active
\citep{mcnamara07}. Multi-wavelength studies show that AGN jet
activity deposits large amounts of mechanical and CR energy into the
host environments, significantly affecting the evolution of gas and
host galaxies. But in our Galaxy there has been little evidence for a
currently-active AGN jet. It would be very unusual if Sgr A* has always been quiescent in the past, 
as during its growth, a large amount of energy 
($\sim 0.1M_{\text BH}c^{2}\sim 10^{60}$ erg) may have been released.
The detection of \emph{Fermi} bubbles as the remnant of
a recent powerful AGN jet event is thus of dramatic astronomical
importance, suggesting that AGN jet activity may also happen regularly
in our Galaxy. 

Furthermore, AGN activity from the GC may contribute significantly to,
or dominate, the CR population in the Galactic halo. Suppose for
example that new pairs of \emph{Fermi} bubbles with total CR energy
$2.7\times 10^{56}$ ergs (twice that of the single bubble in run A1)
are produced every $30$ Myr (AGN duty cycle $\sim 1\%$). This would
supply CR energy to the Galactic halo at a rate $2.86\times 10^{41}$
ergs s$^{-1}$. By comparison, if one supernova of energy $10^{51}$
ergs occurs in the Galaxy every 50 years and 10\% of that energy is
converted to CRs, the total CR power generated, $0.6\times 10^{41}$
ergs s$^{-1}$, is considerably less than the estimated rate from AGN
activity. Nevertheless, CRs from previous generations of \emph{Fermi} bubbles
may not contribute significantly to the locally observed CR energy
density -- this would require that CRs return to the solar vicinity by
relatively slow diffusion in the halo before buoyancy and kinetic
energy of the moving bubbles carry them entirely out of the Galactic
halo.

Do similar \emph{Fermi} bubbles exist in other galaxies?  Due to the limited
sensitivity and resolution of gamma-ray observations, the {\it Fermi}
telescope can not easily detect extragalactic \emph{Fermi} bubbles similar to
those in the Milky Way.  However synchrotron emission of many bubbles
has been detected and spatially resolved in radio observations. AGN
jets have been observed in many spiral Seyfert galaxies where extended
kpc-scale radio structures (KSRs) are common \citep{gallimore06}. In a
sample of 12 Seyfert KSRs, \citet{baum93} found that the extended
radio emission tends to align with the projected minor axis of the
host galaxy, similar to the orientation of the \emph{Fermi} bubbles with
respect to the Galactic disk (see \citealt{elmouttie98} and
\citealt{kharb06} for two well-studied examples). It remains to be
determined if the \emph{Fermi} bubbles and extragalactic KSRs are indeed
similar astrophysical phenomena.

Finally, as in Sections 3.1 and 3.5, we draw attention to two shortcomings of
our CR+hydrodynamic computations of the \emph{Fermi} bubble evolution:
surface irregularities and non-uniform gamma ray limb darkening.
Irregularities in the bubble boundaries are induced by instabilities
in the differentially shearing flow between vertically downflowing gas
in the bubbles and the (assumed) initially stationary gas in the
Galactic halo.  In addition, the rather uniform distribution of cosmic
ray energy density inside our computed bubbles would produce IC
gamma ray intensities that decrease toward the bubble
boundaries, assuming that the ISRF photons (which are upscattered)
are smoothly distributed. These inadequacies may suggest that a new physical mechanism
in addition to the basic jet hydrodynamics, such as shear viscosity discussed in the companion paper -- Paper II, plays a role during the evolution of the \emph{Fermi} bubbles

\acknowledgements 
FG thanks Fabrizio Brighenti, Gregory Dobler, Matthew
McQuinn, S. Peng Oh and Aristotle Socrates for helpful discussions. 
We thank the anonymous referee for an insightful report.
Studies of AGN feedback and the \emph{Fermi} bubbles at UC Santa Cruz are 
supported by NSF and NASA grants for which we are very grateful.

%\bibliography{ms} 

%\bsp 
\label{lastpage}

\end{document}